\documentclass[dvips]{article}
\usepackage{supertabular,lscape,epsfig}
\usepackage{amssymb}
\usepackage{amsmath}
\DeclareSymbolFont{ppa}{OT1}{ppl}{m}{it}
\DeclareMathSymbol{\vv}{\mathalpha}{ppa}{'166}

\begin{document}

\newcommand{\TabCapp}[2]{\begin{center}\parbox[t]{#1}{\centerline{
  \small {\spaceskip 2pt plus 1pt minus 1pt T a b l e}
  \refstepcounter{table}\thetable}
  \vskip2mm
  \centerline{\footnotesize #2}}
  \vskip3mm
\end{center}}

\newcommand{\TTabCap}[3]{\begin{center}\parbox[t]{#1}{\centerline{
  \small {\spaceskip 2pt plus 1pt minus 1pt T a b l e}
  \refstepcounter{table}\thetable}
  \vskip2mm
  \centerline{\footnotesize #2}
  \centerline{\footnotesize #3}}
  \vskip1mm
\end{center}}

\newcommand{\MakeTableSepp}[4]{\begin{table}[p]\TabCapp{#2}{#3}
  \begin{center} \TableFont \begin{tabular}{#1} #4 
  \end{tabular}\end{center}\end{table}}

\newcommand{\MakeTableee}[4]{\begin{table}[htb]\TabCapp{#2}{#3}
  \begin{center} \TableFont \begin{tabular}{#1} #4
  \end{tabular}\end{center}\end{table}}

\newcommand{\MakeTablee}[5]{\begin{table}[htb]\TTabCap{#2}{#3}{#4}
  \begin{center} \TableFont \begin{tabular}{#1} #5 
  \end{tabular}\end{center}\end{table}}

\newcommand{\ie}{{\it i.e.},\,}
\newcommand{\etal}{{\it et al.\ }}
\newcommand{\eg}{{\it e.g.},\,}
\newcommand{\sst}{\scriptscriptstyle}
\newcommand{\fns}{\footnotesize}
\newcommand{\beq}{\begin{equation}
  \renewcommand{\int}{\intop\limits}
  \renewcommand{\oint}{\ointop\limits}}
\newcommand{\eeq}{\end{equation}}
\newcommand{\beqarr}{\par\begin{minipage}{11cm} \begin{eqnarray*}}
\newcommand{\eeqarr}{\end{eqnarray*} \end{minipage} \hfill
   \stepcounter{equation}{\rm (\theequation)}\vspace{3mm}\linebreak}
\newcommand{\bdm}{\begin{displaymath}
  \renewcommand{\int}{\intop\limits}
  \renewcommand{\oint}{\ointop\limits}}
\newcommand{\edm}{\end{displaymath}}
\newcommand{\trule}{\rule{0pt}{14pt}}
\newcommand{\zdot}{\makebox[0pt][l]{.}}
\newcommand{\up}[1]{\ifmmode^{\rm #1}\else$^{\rm #1}$\fi}
\newcommand{\dn}[1]{\ifmmode_{\rm #1}\else$_{\rm #1}$\fi}
\newcommand{\upd}{\up{d}}
\newcommand{\uph}{\up{h}}
\newcommand{\upm}{\up{m}}
\newcommand{\ups}{\up{s}}
\newcommand{\arcd}{\ifmmode^{\circ}\else$^{\circ}$\fi}
\newcommand{\arcm}{\ifmmode{'}\else$'$\fi}
\newcommand{\arcs}{\ifmmode{''}\else$''$\fi}

\newcommand{\Title}[1]{{\large\bf\boldmath #1 
\\[3mm]}}

\newcommand{\Author}[2]{{\large\spaceskip 2pt plus 1pt minus 1pt #1}\\[3mm]
   {\small #2}\\[6mm]}

\newcommand{\Received}[1]{\small\it{ Received #1 }\\}

\newcommand{\Abstract}[2]{{\footnotesize\begin{center}ABSTRACT\end{center}
\vspace{1mm}\par#1\par
\noindent
{\bf Key words:~~}{\it{#2}}}}

\newcommand{\FigCap}[1]{\footnotesize\par\noindent Fig.\  %
  \refstepcounter{figure}\thefigure. #1\par}

\newcommand{\TabCap}[2]{\begin{center}\parbox[t]{#1}{\begin{center}
  \small {\spaceskip 2pt plus 1pt minus 1pt T a b l e}
  \refstepcounter{table}\thetable \\[2mm]
  \footnotesize #2 \end{center}}\end{center}}

\newcommand{\Table}[3]{\begin{table}[htb]\TabCap{#2}{#3}
  \vspace{#1}\end{table}}

\newcommand{\TableSep}[2]{\begin{table}[p]\vspace{#1}
\TabCap{#2}\end{table}}

\newcommand{\TableFont}{\footnotesize}
\newcommand{\TableFontIt}{\ttit}
\newcommand{\SetTableFont}[1]{\renewcommand{\TableFont}{#1}}

\newcommand{\MakeTable}[4]{\begin{table}[htb]\TabCap{#2}{#3}
  \begin{center} \TableFont \begin{tabular}{#1} #4
  \end{tabular}\end{center}\end{table}}

\newcommand{\MakeTableSep}[4]{\begin{table}[p]\TabCap{#2}{#3}
  \begin{center} \TableFont \begin{tabular}{#1} #4
  \end{tabular}\end{center}\end{table}}

\newcommand{\Figure}[2]{\begin{figure}[htb]\vspace{#1}
\FigCap{#2}\end{figure}}

\newenvironment{references}%
{
\footnotesize \frenchspacing
\renewcommand{\thesection}{}
\renewcommand{\in}{{\rm in }}
\renewcommand{\AA}{Astron.\ Astrophys.}
\newcommand{\AAS}{Astron.~Astrophys.~Suppl.~Ser.}
\newcommand{\ApJ}{Astrophys.\ J.}
\newcommand{\ApJS}{Astrophys.\ J.~Suppl.~Ser.}
\newcommand{\ApJL}{Astrophys.\ J.~Letters}
\newcommand{\AJ}{Astron.\ J.}
\newcommand{\IBVS}{IBVS}
\newcommand{\PASP}{P.A.S.P.}
\newcommand{\Acta}{Acta Astron.}
\newcommand{\MNRAS}{MNRAS}
\renewcommand{\and}{{\rm and }}
\section{{\rm REFERENCES}}
\sloppy \hyphenpenalty10000
\begin{list}{}{\leftmargin1cm\listparindent-1cm
\itemindent\listparindent\parsep0pt\itemsep0pt}}%
{\end{list}\vspace{2mm}}

\newcommand{\refitem}[5]{\item[]{#1} #2%
\def\REFARG{#3}\ifx\REFARG\TYLDA\else, {\it#3}\fi
\def\REFARG{#4}\ifx\REFARG\TYLDA\else, {\bf#4}\fi
\def\REFARG{#5}\ifx\REFARG\TYLDA\else, {#5}\fi.}

\newcommand{\Section}[1]{\section{#1}}
\newcommand{\Subsection}[1]{\subsection{#1}}
\newcommand{\Acknow}[1]{\par\vspace{5mm}{\bf Acknowledgements.} #1}

\pagestyle{myheadings}

\newfont{\bb}{ptmbi8t at 12pt}
\newfont{\bbb}{cmbxti10}
\newfont{\bbbb}{cmbxti10 at 9pt}
\newcommand{\uprule}{\rule{0pt}{2.5ex}}
\newcommand{\douprule}{\rule[-2ex]{0pt}{4.5ex}}
\newcommand{\dorule}{\rule[-2ex]{0pt}{2ex}}
\def\thefootnote{\fnsymbol{footnote}}

\begin{center}
\Title{The Optical Gravitational Lensing Experiment. \\ Variable Baseline
Microlensing Events in the Galactic Bulge.\footnote{Based on observations 
obtained with the 1.3~m Warsaw telescope at the Las Campanas Observatory 
of the Carnegie Institution of Washington.}}

\Author{ {\L}.~~W~y~r~z~y~k~o~w~s~k~i$^{1,2}$, ~~A.~~U~d~a~l~s~k~i$^1$,
~~S. ~~M~a~o$^3$, \\ ~~M. ~~K~u~b~i~a~k$^1$,
~~M.\,K.~~S~z~y~m~a~{\'n}~s~k~i$^1$, ~~G.~~P~i~e~t~r~z~y~{\'n}~s~k~i$^{1,4}$,
~~I.~~S~o~s~z~y~{\'n}~s~k~i$^{1,4}$ ~~and ~~O.~~S~z~e~w~c~z~y~k$^1$ } {
$^1$Warsaw University Observatory, Al.~Ujazdowskie~4, 00-478~Warszawa,
Poland\\ 
e-mail:(wyrzykow,udalski,mk,msz,pietrzyn,soszynsk,szewczyk)@astrouw.edu.pl\\
$^2$ Institute of Astronomy, University of Cambridge, Madingley~Road,
Cambridge~CB3~0HA,~UK\\ 
$^3$ University of Manchester, Jodrell Bank Observatory, Macclesfield,
Cheshire~SK11~9DL,~UK\\ 
e-mail: smao@jb.man.ac.uk\\ 
$^4$ Universidad de Concepci{\'o}n, Departamento de Fisica, Casilla~160-C,
Concepci{\'o}n,~Chile}
\vspace*{-9pt}
\end{center}
\vspace*{-15pt}
\Abstract{We present the first systematic search for microlensing events
with variability in their baselines using data from the third phase of the
Optical Gravitational Lensing Experiment (OGLE-III). A total of 137
candidates (88 new) was discovered toward the Galactic bulge. Among these,
21 have periodic oscillations in their baselines, 111 are irregular
variables and 5 are potential long period detached eclipsing binaries.
This is about 10\% of the total number of constant baseline events. They
are hence quite common and can be regarded as a new type of exotic events,
which allow the determination of extra parameters of the events. We show
that microlensing of variable stars may allow us to break the degeneracy
between the blending parameter and magnification. We note that in some
cases variability hidden in the baseline due to strong blending may be
revealed in highly magnified events and resemble other exotic microlensing
behavior, including planetary deviation. A new system (VEWS) for detecting
ongoing variable baseline microlensing events is presented.}
{gravitational lensing - Galaxy: center - Stars: oscillations - Catalogs}
\vspace*{-7pt}
\Section{Introduction}
In the 1990's the first generation microlensing surveys detected first
cases of microlensing events (MACHO -- Alcock \etal 1993, OGLE -- Udalski
\etal 1993), as predicted by Paczy{\'n}ski (1986), opening the new field of
modern astrophysics. Since then several thousand events have been
discovered, mostly toward the Galactic bulge and the rate of detection now
is around 600 per year in real-time systems (\eg the Early Warning System
-- EWS\footnote{{\it http://ogle.astrouw.edu.pl/ogle3/ews/ews.html}},
Udalski 2003). One of the main selection criteria applied in events' search
algorithms was a requirement of constant baseline light curve before and
after the brightening episode. This was used to minimize the risk of a
mistake and interpreting a variable star as a microlensing event. Some
exploding stars which may mimic microlensing usually exhibit variable
baseline. Unfortunately this also excluded variable stars in the sample of
stars searched for microlensing. However, variable stars are microlensed
with the same probability as other stars and because variability is present
in about 10\% of the Galactic bulge stars, these events cannot be
neglected. If discovered they would add about 60 events to 600 being
detected every year, which is a significant number, especially for
statistical studies of microlensing in the Galactic bulge, \eg the
determination of the optical depth. So far only a few events with clear
variability in the baseline were detected, mostly by accident, \eg
MACHO-SMC-1 (\eg Udalski \etal 1997b) or BUL1.725 (Mizerski and Bejger
2002).

In this paper we present the results of the first systematic search for
variable baseline events in the OGLE-III Galactic bulge data covering
seasons 2001--2004. We discovered a total of 137 (88 new) candidate events,
among which 21 have periodic oscillations in their baselines, 111 are
irregular variables and 5 are potential long period detached eclipsing
binaries. We present simple models of periodic baseline microlensing
events, analyze the behavior of the amplitude of variability during
gravitational magnification and show that the change of amplitude of
brightness variation during the microlensing allows a direct determination
of the blending fraction in some events. This method was recently
successfully applied to the event MACHO 97-SMC-1 (Assef \etal 2006), which
exhibits periodic variability in its baseline. We show that these events
are common and can be regarded as a new type of exotic microlensing events,
from which additional information (\eg distance to the source or lens) can
be extracted when combined with further spectroscopic observations. We
conclude the paper by presenting an extension to the OGLE Early Warning
System (EWS), which can detect microlensing episodes of variable stars in
real-time and allows frequent followup in order to take full advantage of
this type of events.

\Section{Model of Microlensing Event with Variable Baseline}
Whenever the variability in the baseline is regular and predictable in any
moment of time it is possible to find a good model of the microlensing
event of a variable star. As ground-based microlensing surveys always deal
with very crowded fields, observed star-like objects are usually a
composite of several individual stars. Variability of at least one of these
blended stars will be observed as a change of brightness in the whole
composite object. During a microlensing event light of one of the stars is
magnified. Thus, there are two classes of such microlensing event models
with variable baseline, depending on whether the variable component is
magnified or just acts as a blend. For brevity, we refer to the first as a
``variable source'' event and the second as a ``variable blend'' event.

If $I_{\rm var}(t)$ is an analytical or approximated function of the
baseline magnitude variability (\eg from Fourier series fit or
interpolation between phase-folded baseline data points), the addition to
the minimum brightness flux due to variability equals to:
\vspace*{-5pt}
$$f_{\rm var}(t)=10^{0.4|I_0-I_{\rm var}(t)|}-1\eqno(1)$$
where $I_0$ is in this case the minimum magnitude of the variable baseline
(see Fig.~2 for an illustration). In the case where the variable source is
microlensed, the flux relative to the minimum baseline value is given by:
\vspace*{-5pt}
$$f(t)=\Big(f_s+f_{\rm var}(t)\Big)A(t)+f_b.\eqno(2)$$
In the case where the variability comes from a blended, unlensed
variable star, we have
\vspace*{-6pt}
$$f(t)=f_sA(t)+\Big(f_b+f_{\rm var}(t)\Big).\eqno(3)$$
In Eqs.(2)--(3) $f_s$ denotes the fraction of the total flux contributed by
the lensed source to the composite at the minimum brightness of the
baseline, and $f_b(\equiv1-f_s)$ is the fraction of light contributed by
any additional blend.

$A(t)$ is the microlensing magnification, in the simplest case of single
point mass lens given in the standard form (Paczy{\'n}ski 1986):
\vspace*{-6pt}
$$A(t)=\frac{u(t)^2+2}{u(t)\sqrt{u(t)^2+4}},\qquad
u(t)=\sqrt{u_0^2+\tau(t)^2},\qquad
\tau(t)=\frac{t-t_0}{t_E}\eqno(4)$$
where $u_0$ is the minimum impact parameter in units of the Einstein radius
at the moment $t_0$, and $t_E$ is the Einstein radius crossing time.

\Section{Variability Amplitude in the Microlensing Event}
Using the above microlensing model, we can examine the behavior of the
variability amplitude during a microlensing event involving a variable
star. In the simplest case of a single variable star, its amplitude of
variability ($\Delta I_A$) does not change with amplification $A$ and
remains equal to the amplitude in the baseline ($\Delta I_{\rm base}$), as
the ratio of the amplified to the baseline flux variations is constant.

However, when the lensed variable star is blended with some additional
constant flux, the observed amplitude will increase with $A$ as:
\vspace*{-6pt}
$$\Delta I_A=2.5\log\left[1+\frac{\Delta fA}{f_sA+f_b}\right]\eqno(5)$$
\vspace*{-2pt}
where $\Delta f=10^{0.4\Delta I_{\rm base}}-1$ is the baseline variability
amplitude.

\newpage
In the theoretical limit of infinite amplification the observed amplitude
approaches some constant value, which is the amplitude of the variability
of the star we would measure if there were no blending. This real amplitude
equals to $2.5\log\left(1+\frac{\Delta f}{f_s}\right)$ and its dependence
on the blending and baseline variability amplitude is shown in Fig.~1.
\begin{figure}[htb]
\includegraphics[width=12.5cm]{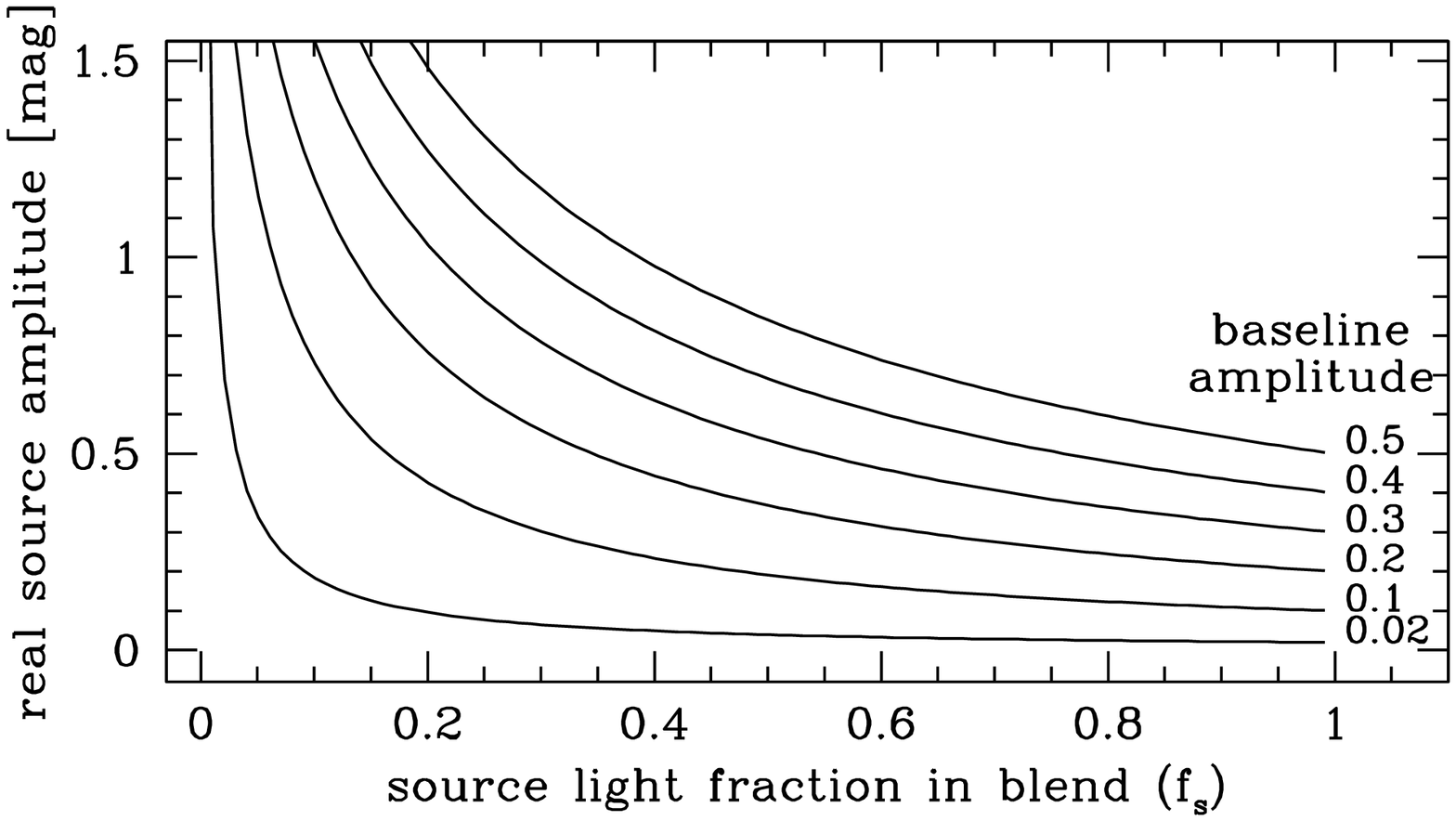} 
\FigCap{The dependence of the real amplitude of variability, $\Delta I_A$,
on the blending parameter, $f_s$. The baseline variability amplitude,
$\Delta I_{\rm base}$, is indicated for each curve.}
\end{figure}

This implies that in the case of very strong blending and/or small
intrinsic amplitude the variability in the baseline may be buried in the
photometry noise and undetectable, making the baseline resembling that of a
constant star. However, the hidden variability may then be revealed thanks
to microlensing and it will show deviations in the light curve from the
standard shape, especially near the peak of magnification. This effect
should be considered especially in the case of high magnification events in
the crowded Galactic bulge fields, as the variation of brightness due to
lensed source variability may be mistaken with other microlensing effects,
\eg finite source size, parallax or planetary deviation.

As the observed amplification ($A_{\rm obs}$) in a blended event is not the
real microlensing amplification, the latter is related to the blending by:
$$A_{\rm unblended}=\frac{A_{\rm obs}-1}{f_s}+1.\eqno(6)$$

A simple calculation of flux changes during microlensing of a variable
source star using Eq.~(6) leads to:
$$f_s=\Delta f\frac{F_1-1}{F_2-F_1-\Delta f},\quad
F_1=10^{0.4D_{\rm mag}},\quad
F_2=10^{0.4(D_{\rm mag}+\Delta I_A)}\eqno(7)$$
$D_{\rm mag}$ is the magnification in magnitudes, measured along the bottom
of microlensed variable light curve from $I_0$, and $\Delta I_A$ is the
amplitude expected at that magnification. Fig.~2 shows modeled variable
baseline light curve with $D_{\rm mag}$, $\Delta I_A$, $I_0$ and $\Delta
I_{\rm base}$ marked. Eq.~(7) is a simple formula, which allows one to
derive the blending parameter directly from the light curve using its
quantities. The accuracy of this formula is limited by the accuracy with
which we can measure the variability amplitude during the event. It can be
used to complement the usual value obtained from fitting the data with the
model of a variable microlensed source (Eq.~2), or in the case of well
sampled detached eclipsing binaries even as a substitute. Fig.~3 shows an
example diagram for determining the blending parameter for an event with a
baseline variability amplitude of 0.2~mag.
\begin{figure}[htb]
\includegraphics[width=13cm]{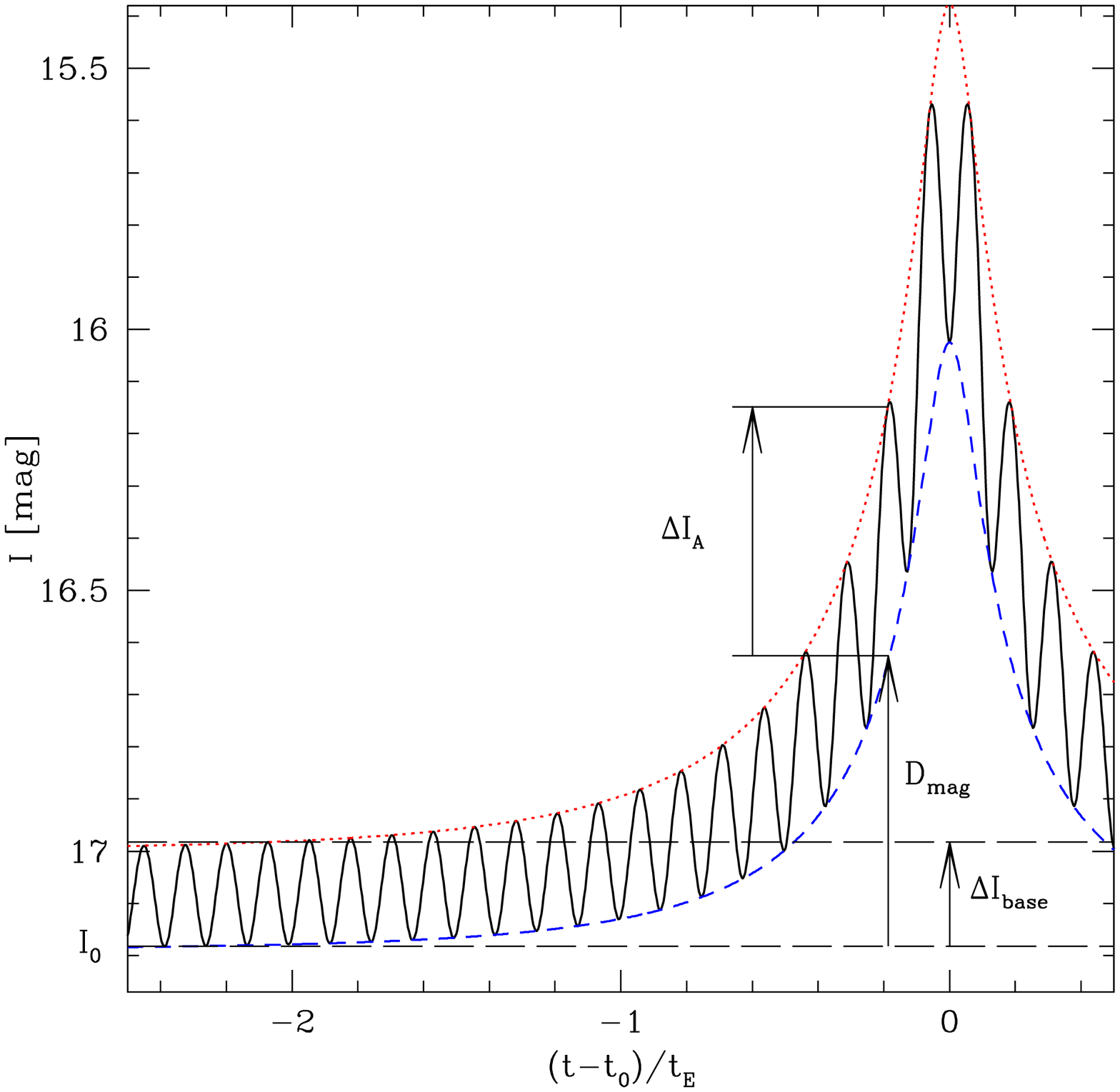}
\FigCap{Synthetic light curve of a variable source microlensing event. The
dotted and short-dashed lines outline the amplitude of brightness
variability. The bottom long-dashed line represents the minimum baseline
magnitude $I_0$. $D_{\rm mag}$, $\Delta I_{\rm base}$ and $\Delta I_A$
from Eq.~(7) are labeled.}
\end{figure}
\begin{figure}[htb]
\includegraphics[width=12.5cm]{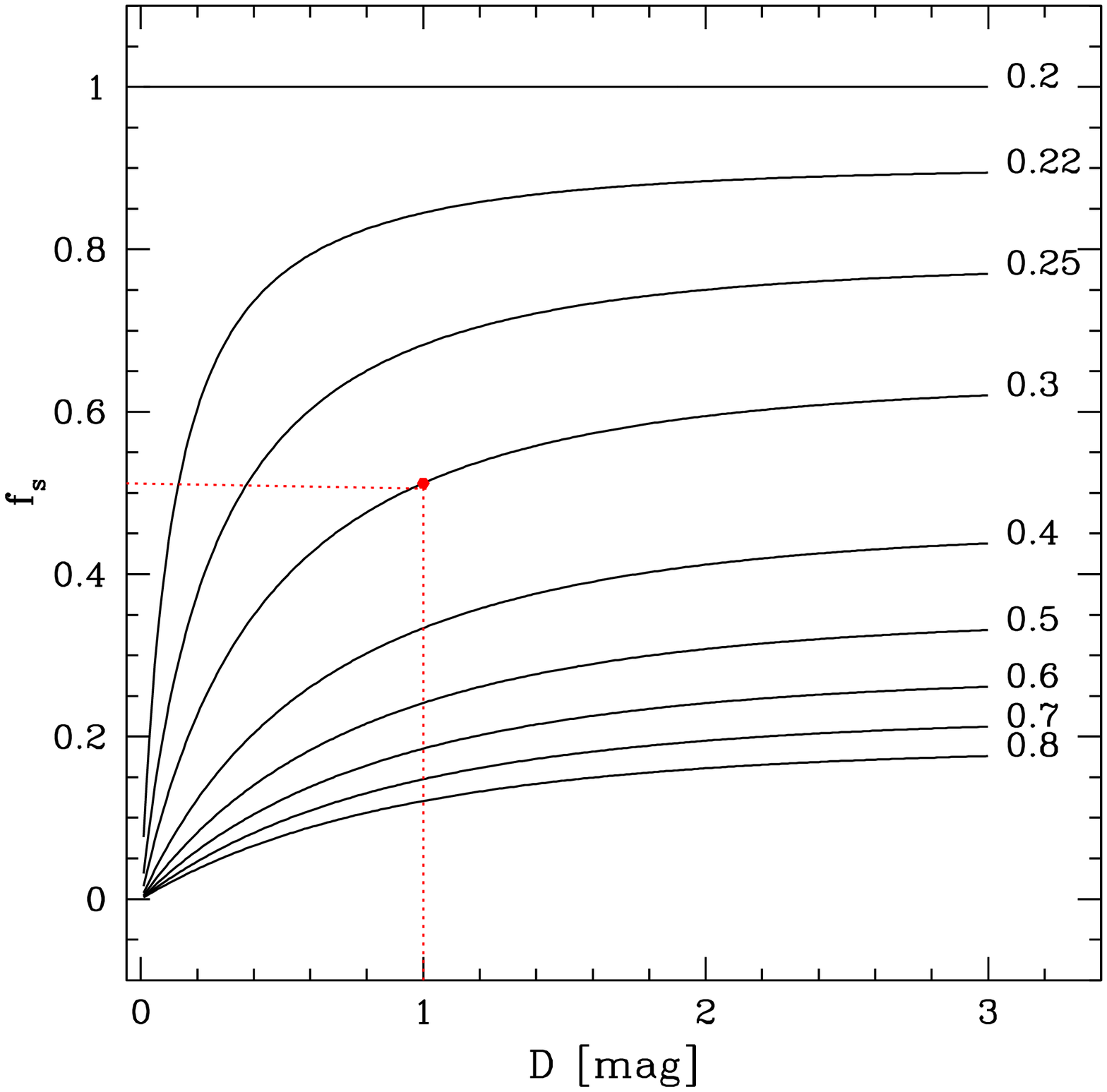}
\FigCap{Example plot for determining the blending parameter from the
microlensing light curve of a variable source event with baseline amplitude
0.2~mag. The dashed lines show how to determine $f_s$ if the amplified
amplitude equals $\Delta I_A=0.3$~mag on top of the brightening of $D_{\rm
mag}=1$~mag. $f_s = 0.51$ for this example.}
\end{figure}

When the variability observed in the composite object comes from one of the
blending stars and the microlensed star is constant, the observed amplitude
will decrease with amplification:
$$\Delta I_A=2.5\log\left[1+\frac{\Delta f}{f_s A+f_b}\right].\eqno(8)$$
With $A$ increasing to infinity the amplitude declines and vanishes
completely at the peak.

In the case of a variable blend, because of Eq.~(6), it is not possible to
obtain a similar relation to Eq.~(7) and the only way to measure the
blending fraction in such a microlensing event is through fitting the data
with a model (Eqs.~3 and 4).

\renewcommand{\arraystretch}{1.1}
\MakeTableSep{ccc|ccc}{12cm}{Equatorial coordinates (J2000) of centers of the selected
82 OGLE-III fields}
{\hline
\noalign{\vskip4pt}
field & $\alpha_{2000}$ & $\delta_{2000}$ & field & $\alpha_{2000}$ & $\delta_{2000}$\\
\noalign{\vskip4pt}
\hline
\noalign{\vskip4pt}
BLG100 & 17\uph51\upm00\zdot\ups43 & $-30\arcd04\arcm21\zdot\arcs6$ & 
BLG175 & 18\uph03\upm28\zdot\ups19 & $-31\arcd19\arcm22\zdot\arcs1$ \\
BLG101 & 17\uph53\upm40\zdot\ups61 & $-29\arcd54\arcm25\zdot\arcs4$ & 
BLG176 & 18\uph06\upm09\zdot\ups29 & $-31\arcd19\arcm22\zdot\arcs3$  \\
BLG102 & 17\uph56\upm20\zdot\ups64 & $-29\arcd35\arcm24\zdot\arcs8$ & 
BLG179 & 17\uph50\upm00\zdot\ups36 & $-30\arcd43\arcm50\zdot\arcs7$   \\
BLG103 & 17\uph56\upm20\zdot\ups60 & $-30\arcd10\arcm55\zdot\arcs6$ & 
BLG180 & 17\uph52\upm40\zdot\ups25 & $-30\arcd43\arcm53\zdot\arcs2$   \\
BLG104 & 17\uph59\upm00\zdot\ups66 & $-29\arcd32\arcm50\zdot\arcs2$ &
BLG181 & 17\uph55\upm20\zdot\ups11 & $-30\arcd43\arcm51\zdot\arcs3$   \\
BLG105 & 18\uph01\upm40\zdot\ups03 & $-29\arcd32\arcm45\zdot\arcs9$ & 
BLG182 & 17\uph58\upm00\zdot\ups48 & $-30\arcd43\arcm50\zdot\arcs6$   \\
BLG108 & 17\uph46\upm29\zdot\ups73 & $-36\arcd38\arcm43\zdot\arcs0$ & 
BLG183 & 18\uph00\upm40\zdot\ups04 & $-30\arcd43\arcm50\zdot\arcs2$   \\
BLG115 & 17\uph54\upm59\zdot\ups54 & $-36\arcd03\arcm19\zdot\arcs5$ & 
BLG184 & 18\uph03\upm18\zdot\ups41 & $-30\arcd43\arcm53\zdot\arcs3$   \\
BLG117 & 17\uph49\upm18\zdot\ups11 & $-35\arcd27\arcm49\zdot\arcs4$ & 
BLG185 & 18\uph06\upm01\zdot\ups00 & $-30\arcd43\arcm53\zdot\arcs1$   \\
BLG119 & 17\uph54\upm56\zdot\ups17 & $-35\arcd27\arcm50\zdot\arcs6$ & 
BLG188 & 17\uph59\upm00\zdot\ups64 & $-30\arcd08\arcm32\zdot\arcs4$   \\
BLG121 & 17\uph46\upm28\zdot\ups63 & $-34\arcd52\arcm16\zdot\arcs8$ & 
BLG189 & 18\uph01\upm39\zdot\ups48 & $-30\arcd08\arcm30\zdot\arcs4$   \\
BLG122 & 17\uph49\upm17\zdot\ups69 & $-34\arcd52\arcm23\zdot\arcs9$ & 
BLG190 & 18\uph04\upm28\zdot\ups47 & $-30\arcd08\arcm30\zdot\arcs0$   \\
BLG129 & 17\uph43\upm43\zdot\ups64 & $-34\arcd16\arcm44\zdot\arcs8$ & 
BLG193 & 18\uph12\upm25\zdot\ups41 & $-30\arcd08\arcm30\zdot\arcs4$   \\
BLG130 & 17\uph46\upm30\zdot\ups76 & $-34\arcd16\arcm45\zdot\arcs4$ & 
BLG194 & 17\uph51\upm00\zdot\ups21 & $-29\arcd29\arcm01\zdot\arcs2$   \\
BLG131 & 17\uph49\upm17\zdot\ups75 & $-34\arcd16\arcm46\zdot\arcs1$ & 
BLG195 & 17\uph53\upm38\zdot\ups80 & $-29\arcd18\arcm53\zdot\arcs7$   \\
BLG132 & 17\uph52\upm04\zdot\ups73 & $-34\arcd16\arcm47\zdot\arcs5$ & 
BLG196 & 18\uph04\upm27\zdot\ups69 & $-29\arcd32\arcm40\zdot\arcs2$   \\
BLG133 & 17\uph54\upm51\zdot\ups74 & $-34\arcd16\arcm47\zdot\arcs9$ & 
BLG197 & 18\uph07\upm04\zdot\ups75 & $-29\arcd32\arcm55\zdot\arcs4$   \\
BLG134 & 17\uph57\upm38\zdot\ups65 & $-34\arcd16\arcm47\zdot\arcs4$ & 
BLG198 & 18\uph09\upm42\zdot\ups68 & $-29\arcd32\arcm55\zdot\arcs5$   \\
BLG138 & 17\uph45\upm15\zdot\ups24 & $-33\arcd41\arcm14\zdot\arcs3$ & 
BLG205 & 17\uph57\upm16\zdot\ups84 & $-28\arcd57\arcm43\zdot\arcs6$   \\
BLG139 & 17\uph48\upm00\zdot\ups12 & $-33\arcd41\arcm16\zdot\arcs0$ & 
BLG206 & 17\uph59\upm53\zdot\ups28 & $-28\arcd57\arcm33\zdot\arcs3$   \\
BLG140 & 17\uph50\upm45\zdot\ups18 & $-33\arcd41\arcm25\zdot\arcs2$ & 
BLG208 & 18\uph05\upm08\zdot\ups83 & $-28\arcd57\arcm33\zdot\arcs4$   \\
BLG141 & 17\uph53\upm30\zdot\ups27 & $-33\arcd41\arcm20\zdot\arcs4$ & 
BLG219 & 18\uph10\upm30\zdot\ups46 & $-28\arcd22\arcm05\zdot\arcs3$  \\
BLG142 & 17\uph56\upm15\zdot\ups58 & $-33\arcd41\arcm36\zdot\arcs9$ & 
BLG220 & 18\uph13\upm06\zdot\ups43 & $-28\arcd22\arcm05\zdot\arcs5$ \\
BLG147 & 17\uph49\upm45\zdot\ups43 & $-33\arcd06\arcm04\zdot\arcs5$ & 
BLG225 & 18\uph04\upm30\zdot\ups63 & $-27\arcd46\arcm31\zdot\arcs6$   \\
BLG148 & 17\uph52\upm36\zdot\ups75 & $-33\arcd06\arcm05\zdot\arcs8$ & 
BLG226 & 18\uph07\upm05\zdot\ups75 & $-27\arcd46\arcm32\zdot\arcs6$   \\
BLG149 & 17\uph55\upm15\zdot\ups67 & $-33\arcd06\arcm06\zdot\arcs7$ & 
BLG227 & 18\uph09\upm39\zdot\ups95 & $-27\arcd46\arcm32\zdot\arcs3$   \\
BLG150 & 17\uph58\upm00\zdot\ups67 & $-33\arcd06\arcm06\zdot\arcs6$ & 
BLG232 & 18\uph00\upm00\zdot\ups07 & $-27\arcd10\arcm59\zdot\arcs1$   \\
BLG155 & 17\uph52\upm18\zdot\ups04 & $-32\arcd30\arcm32\zdot\arcs3$ & 
BLG236 & 18\uph10\upm20\zdot\ups52 & $-27\arcd11\arcm01\zdot\arcs8$  \\
BLG156 & 17\uph55\upm01\zdot\ups07 & $-32\arcd30\arcm33\zdot\arcs7$ & 
BLG249 & 18\uph06\upm00\zdot\ups55 & $-25\arcd59\arcm51\zdot\arcs6$   \\
BLG157 & 17\uph57\upm44\zdot\ups20 & $-32\arcd30\arcm33\zdot\arcs8$ & 
BLG250 & 18\uph08\upm32\zdot\ups97 & $-26\arcd00\arcm00\zdot\arcs2$   \\
BLG158 & 18\uph00\upm27\zdot\ups14 & $-32\arcd30\arcm35\zdot\arcs8$ & 
BLG251 & 18\uph11\upm07\zdot\ups91 & $-26\arcd00\arcm01\zdot\arcs0$   \\
BLG160 & 18\uph05\upm53\zdot\ups28 & $-32\arcd30\arcm34\zdot\arcs9$ & 
BLG252 & 18\uph13\upm39\zdot\ups54 & $-25\arcd59\arcm53\zdot\arcs7$   \\
BLG163 & 17\uph52\upm45\zdot\ups08 & $-31\arcd54\arcm46\zdot\arcs3$ & 
BLG333 & 17\uph35\upm30\zdot\ups51 & $-27\arcd14\arcm34\zdot\arcs2$   \\
BLG164 & 17\uph55\upm27\zdot\ups07 & $-31\arcd54\arcm45\zdot\arcs9$ & 
BLG339 & 17\uph47\upm00\zdot\ups35 & $-22\arcd34\arcm33\zdot\arcs0$  \\
BLG165 & 17\uph58\upm08\zdot\ups87 & $-31\arcd54\arcm46\zdot\arcs8$ & 
BLG340 & 17\uph49\upm30\zdot\ups56 & $-22\arcd34\arcm27\zdot\arcs4$   \\
BLG166 & 18\uph00\upm51\zdot\ups02 & $-31\arcd54\arcm47\zdot\arcs8$ & 
BLG342 & 17\uph46\upm30\zdot\ups11 & $-23\arcd09\arcm56\zdot\arcs1$   \\
BLG167 & 18\uph03\upm32\zdot\ups99 & $-31\arcd54\arcm48\zdot\arcs5$ &
BLG343 & 17\uph49\upm00\zdot\ups20 & $-23\arcd09\arcm55\zdot\arcs1$   \\
BLG171 & 17\uph52\upm44\zdot\ups51 & $-31\arcd19\arcm20\zdot\arcs6$ & 
BLG346 & 17\uph42\upm00\zdot\ups67 & $-24\arcd21\arcm02\zdot\arcs6$  \\
BLG172 & 17\uph55\upm25\zdot\ups46 & $-31\arcd19\arcm20\zdot\arcs6$ & 
BLG347 & 17\uph44\upm31\zdot\ups74 & $-24\arcd20\arcm58\zdot\arcs8$   \\
BLG173 & 17\uph58\upm06\zdot\ups41 & $-31\arcd19\arcm21\zdot\arcs3$ & 
BLG352 & 17\uph39\upm20\zdot\ups60 & $-21\arcd09\arcm30\zdot\arcs0$   \\
BLG174 & 18\uph00\upm47\zdot\ups30 & $-31\arcd19\arcm21\zdot\arcs6$ & 
BLG354 & 17\uph35\upm01\zdot\ups74 & $-23\arcd39\arcm32\zdot\arcs1$  \\
\hline}

\Section{Observational Data}
\vspace*{9pt}
All photometric data presented in this paper were collected with the 1.3-m
Warsaw Telescope at the Las Campanas Observatory, Chile, which is operated
by the Carnegie Institution of Washington, during the third phase of the
OGLE project. OGLE-III CCD mosaic camera consists of 8 CCD detectors, $2048
\times4096$ pixels of 15 $\mu$m size each. Every pixel corresponds to a
0\zdot\arcs26 scale, thus the whole $8192\times8192$ pixel mosaic gives a
field of view of $35\arcm\times35\arcm$. More details about the telescope
and the instrumentation can be found in Udalski, Kubiak and Szyma{\'n}ski
(1997a) and Udalski (2003).

All images collected by the OGLE-III telescope were de-biased and
flat-fielded with the automatic software almost in real-time. Then, they
were processed by the photometric data pipeline (Udalski 2003) using the
image subtraction method, based on the Wo{\'z}niak's (2000) implementation
of the Difference Image Analysis (DIA) technique (Alard and Lupton
1998). Finally, all photometric measurements for all stars in given field
were stored in the OGLE database (\eg Szyma{\'n}ski and Udalski 1993),
separately for each filter and chip of the mosaic.

Since its beginning in 2001 OGLE-III observed about 300 fields with
different frequency. Among them we selected 82 which were monitored
frequently in seasons 2001--2004, \ie had at least 150 data points in the
{\it I}-band, in which the vast majority of the observations were taken.
The total number of stars brighter than 21 mag (the rough OGLE telescope
limit), in all 82 fields is about 100 million, varying from 800 thousand to
2 million per field. Table~1 presents equatorial coordinates of all the
selected fields.

It is commonly known that the errors obtained with the DIA photometry are
generally under-estimated (\eg Wo{\'z}niak 2000), especially for bright
stars. Alard and Lupton (1998) attribute photometry scatters larger than
the Poisson noise to significant movement of the centroid of the star due
to atmospheric effects. Various solutions of this problem for the OGLE-III
microlensing events were applied up-to-date, \eg Snodgrass \etal (2004),
Collinge (2004), but in general they assumed constant baseline and thus
these methods cannot be applied to variable baseline events. Therefore, we
applied the technique used previously for OGLE-I photometry (Udalski \etal
1994), which was an application of an earlier algorithm of Lupton \etal
(1989). The method uses constant stars, separately in each field, to obtain
correction factors to error bars for a given magnitude and the original
error. Correction factors determined for {\it I}-band magnitudes from 15 to
18 range from 1.1 to 1.4, depending on the field. However, for bright
measurements ($I\le15$~mag) corrections increase with increasing
brightness, and can be as large as 2 to 3. Error bars of faint measurements
(about 19--20~mag) were slightly over-estimated and had to be multiplied by
0.5~--~0.8.

All further analyzes were performed using rescaled photometry errors. Note
that all magnitude measurements are only roughly calibrated, but as this
only affects the zero-point of the light curve, it has little impact on our
variability and microlensing studies.

\Section{Search Algorithm}
We searched for microlensing events with variable baseline among all
objects in the database for 82 selected OGLE-III Bulge fields. We analyzed
each chip of each field separately; each chip contains from 60 to 270
thousand stars, depending on the stellar density.

First, from the error-corrected light curves we removed data points more
than three standard deviations from the mean brightness of the whole light
curve provided that their nearest neighboring points deviate by less than
two standard deviations. Then, every object from the database exhibiting
any kind of variation was investigated for the presence of a significant
brightening peak. At this stage we applied the algorithm used in the search
for microlensing events in the OGLE-II data as described in details in Sumi
\etal (2006). The algorithm calculated the significance $\sigma_i$ of each
data point with respect to the baseline within a continuous running window,
A, spanning half of the total time span of all measurements:
$$\sigma_i=\frac{I_{{\rm med},B}-I_i}{\sqrt{{\Delta I_i}^2+{\sigma_{B}}^2}}\eqno(9)$$
where $I_i$ and $\Delta I_i$ are the magnitude and error of the {\it i}th
data point, respectively, and $I_{{\rm med},B}$ and $\sigma_{B}$ are,
respectively, the median and standard deviation of the magnitude in window
B, which is outside A. Fig.~4 schematically shows the idea of the search
procedure. This procedure can detect events with both constant and small
amplitude varying baselines, but in order to allow for larger amplitudes of
variability we masked out all data points below the median of all magnitude
measurements, as shown in Fig.~4.
\begin{figure}[htb]
\includegraphics[width=12.5cm]{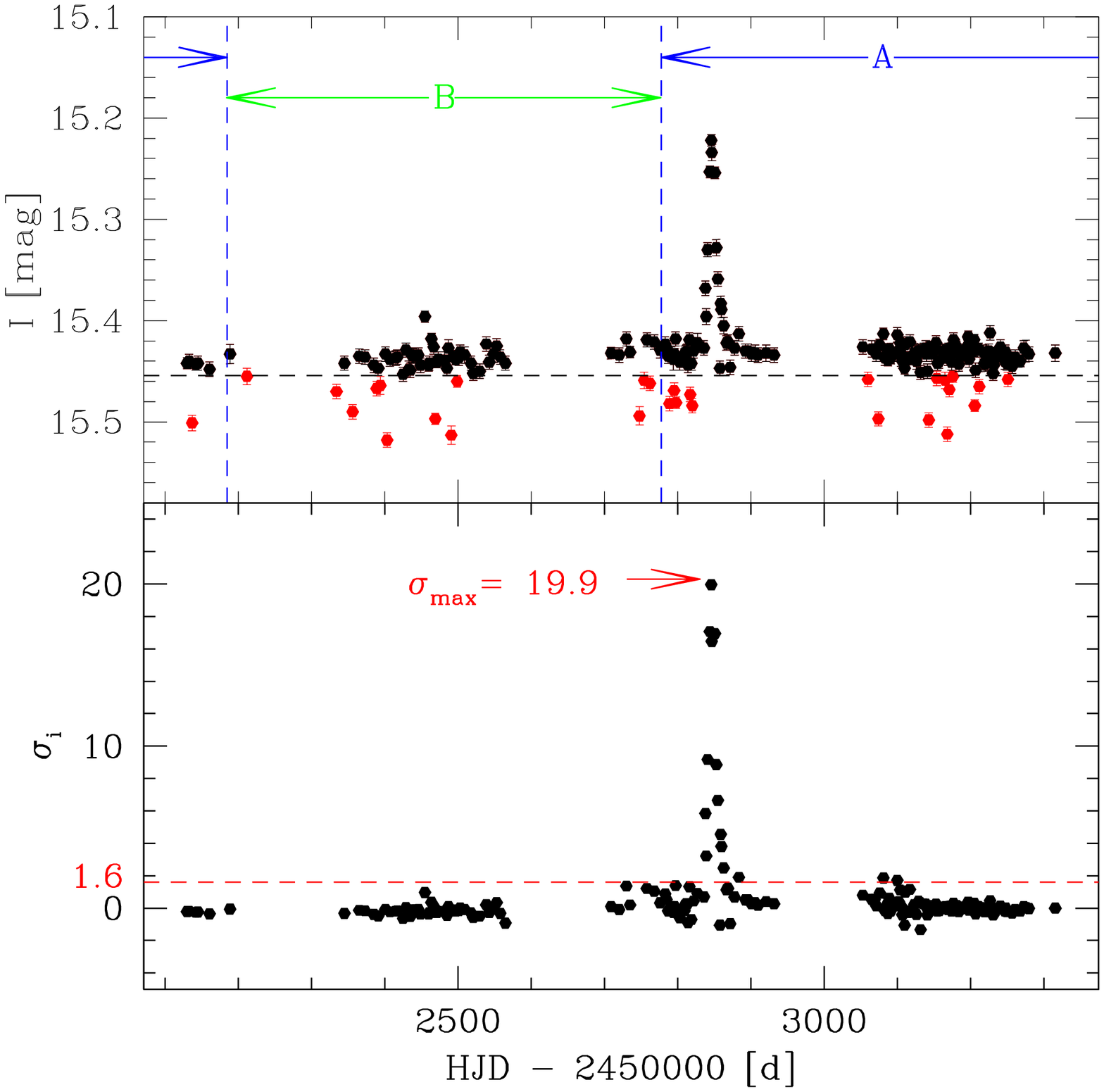}
\FigCap{Visualization of the search procedure. The upper panel shows the
light curve of an event with eclipsing variability in the baseline. Points
below the median (horizontal dashed line) were excluded from significance
calculations. The significance of the central point of the window A was
calculated with respect to the data in window B. The lower panel shows the
significance calculated for each data point. The minimum detection
threshold of 1.6 and the maximum significance value for this event are
marked.}
\end{figure}

In the next step for all the candidate light curves, we removed the
brightening episode and checked the remaining data for periodicity in
the range of 0.25 to 500 days. All the candidates were then inspected
visually and as a result 21 events with periodic baselines were selected
together with 111 candidates with irregular baselines. In addition, we
found five events with a few data points significantly below the constant
baseline, which can potentially be caused by eclipsing variability. Tables
2 and 4 contain information for the periodic and irregular variables
respectively. The name of each event is created with the same convention as
used in previous OGLE catalogs (\eg {\.Z}ebru{\'n} \etal 2001, Wyrzykowski
\etal 2003) by joining the right ascension and declination, for example,
the coordinates of OGLE180047.11--285934.5 are
$\alpha_{2000}=18\uph00\upm47\zdot\ups11$ and
$\delta_{2000}=-28\arcd59\arcm34\zdot\arcs5$ (J2000). If an event was
cross-identified with an object in the EWS, the latter designation is also
given.

\MakeTable{lcccc}{12.5cm}{Microlensing event candidates with periodic
baseline}
{\hline
\multicolumn{1}{c}{name} & period & $T_0$  & $I_0$   & remarks \\
     & [d] & [HJD-245000] & [mag] &  (EWS)  \\
\hline
OGLE174448.21-335605.4 & 0.36380 & 2123.73901 & 19.86$\pm$0.18 &
2003-BLG-428 \\
\hline
OGLE174841.52-352427.0 & 2.893016 & 2124.05151 & 18.17$\pm$0.03 & - \\
\hline
OGLE174913.88-334923.7 & 0.69884 & 2798.03955 & 19.69$\pm$0.14 & - \\
\hline
OGLE175239.94-322613.3 & 0.82756 & 2124.72705 & 19.02$\pm$0.17 &
2002-BLG-103 \\
\hline
OGLE175301.80-310449.9 & 0.533903 & 2790.37134 & 17.18$\pm$0.02 & - \\
\hline
OGLE175341.77-323002.2 & 0.644129 & 2125.40332 & 19.67$\pm$0.20 & - \\
\hline
OGLE175535.13-331533.3 & 7.13057 & 3512.74393 & 16.81$\pm$0.01 & - \\
\hline
OGLE175554.26-301826.7 & 4.026575 & 2112.56177 & 17.07$\pm$0.01 &
2003-BLG-452 \\
\hline
OGLE175646.40-304432.7 & 0.54399 & 2129.36426 & 17.83$\pm$0.06 & - \\
\hline
OGLE175708.46-302007.1 & 10.636035 & 2107.67065 & 16.71$\pm$0.03 &
$P_2$=1.10047 d$^*$ \\
\hline
OGLE175828.21-304717.4 & 0.34825 & 2128.58350 & 18.00$\pm$0.05 &
2004-BLG-390 \\
\hline
OGLE175907.77-305519.1 & 0.86193 & 2147.69800 & 18.84$\pm$0.09 &
2004-BLG-101 \\
\hline
OGLE180047.11-285934.5 & 3.134806 & 2794.80420 & 15.52$\pm$0.01 & - \\
\hline
OGLE180502.14-293703.4 & 7.584951 & 2132.95898 & 19.98$\pm$0.27 & - \\
\hline
OGLE180503.30-293145.4 & 11.567380 & 2122.79810 & 17.59$\pm$0.06 & - \\
\hline
OGLE180834.52-255044.9 & 1.14626 & 3152.76758 & 19.75$\pm$0.14 & - \\
\hline
OGLE180950.21-260542.7 & 0.62610 & 2129.88306 & 18.93$\pm$0.10 &
2002-BLG-093 \\
\hline
OGLE181203.58-255823.5 & 0.539121 & 2129.54321 & 18.62$\pm$0.06 & - \\
\hline
OGLE181322.09-294846.4 & 53.2198 & 2107.63257 & 17.47$\pm$0.03 &
2003-BLG-139 \\
\hline
OGLE175618.55-294252.9 & 21.02317 & 2794.96635 & 14.174$\pm$0.005 & - \\
\hline
OGLE180540.47-273427.5 & 3.96628 & 2406.8730 & 17.21$\pm$0.03 & 2004-BLG-081 \\
\hline
\multicolumn{5}{l}{$^*$second, weaker period in the baseline} \\
}

\MakeTable{l@{\hspace{-13pt}}rrrrccc}{12.5cm}{Best-fit microlensing model
parameters for three events with periodic baseline}
{\hline
\multicolumn{1}{c}{name} & 
\multicolumn{1}{c}{$t_0$} &
\multicolumn{1}{c}{$t_E$} &
\multicolumn{1}{c}{$u_0$} & 
\multicolumn{1}{c}{$f_s$} & 
\multicolumn{1}{c}{$\chi^2$} & 
\multicolumn{1}{c}{$dof$} & 
\multicolumn{1}{c}{model} \\
     & 
\multicolumn{1}{c}{[HJD-2450000]} & 
\multicolumn{1}{c}{[d]} &
       &       &    & & \\
\hline
OGLE174841.52-352427.0 & 2897.9 & 57.85 & 0.7461 & 0.89 & 642.8 & 258 & var.source \\
& $\pm$0.9 & $\pm$5.12 & $\pm$0.0937 & $\pm$0.19 & & & \\
\hline
OGLE175907.77-305519.1 & 3095.0 & 21.37 & 0.1292 & 0.35 & 449.0 & 280 & var.blend \\
& $\pm$0.1 & $\pm$2.59 & $\pm$0.0211 & $\pm$0.04 & & & \\
\hline
OGLE180047.11-285934.5 & 2847.1 & 6.73 & 0.9567 & 0.57 & 500.4 & 295 & var.source \\
& $\pm$0.1 & $\pm$1.74 & $\pm$0.3726 & $\pm$0.36 & & & \\
 & 2847.1 & 8.33 & 0.6731 & 0.33 & 499.2 & 295 & var.blend \\
& $\pm$0.1 & $\pm$2.09 & $\pm$0.2674 & $\pm$0.17 & & & \\
\hline}

\Section{Events with Periodic Baselines}
Light curves of events and their phase-folded periodic baselines are shown
in Appendix~A and listed in Table~2. The event baselines with periodic
variability were modeled using Fourier harmonics. However, for three
detached eclipsing variables (OGLE174913.88--334923.7,
OGLE175535.13--331533.3 and OGLE180047. 11--285934.5) models were obtained
with the EBAS code (Tamuz \etal 2006). Using a model of each event's
baseline we then fitted each event with a microlensing model as discussed
in Section~2. In the majority of cases both the variable source and
variable blend models did not converge to a unique model, mostly due to a
small number of data points during the event of magnification. The model
was only fitted to the data of three events. Table~3 presents the model
parameters for each event, the value of the $\chi^2$ and the number of
degrees of freedom. The observed light curves of these three events are
shown in Figs.~5--7 together with the models.
\begin{figure}[p]
\vglue-3mm
\centerline{\includegraphics[width=9.3cm]{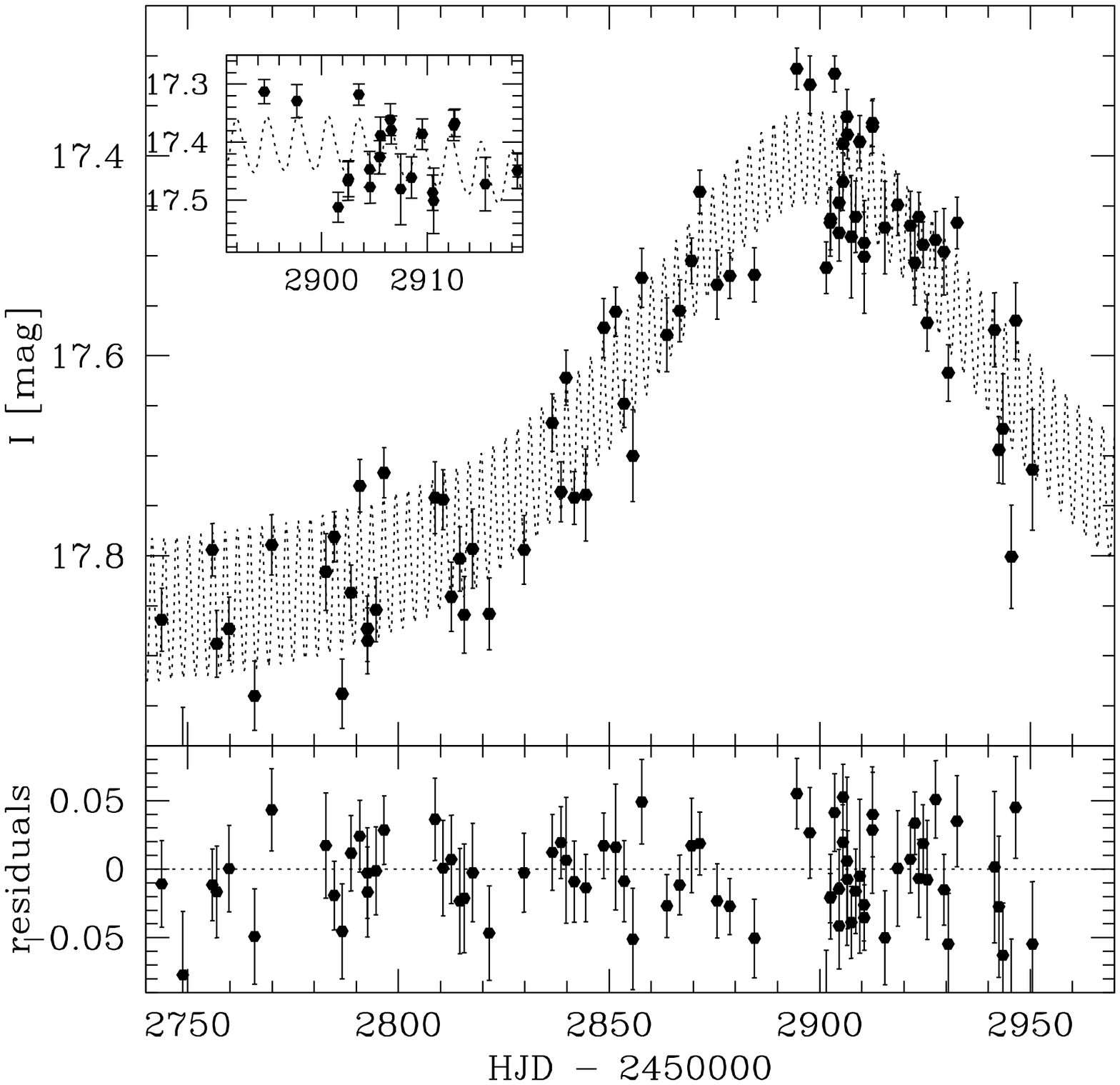}}
\FigCap{Data and best-fit microlensing models with variable baseline of
event OGLE174841.52--352427.0. The dotted line shows the
variable source model.}
\centerline{\includegraphics[width=9.3cm]{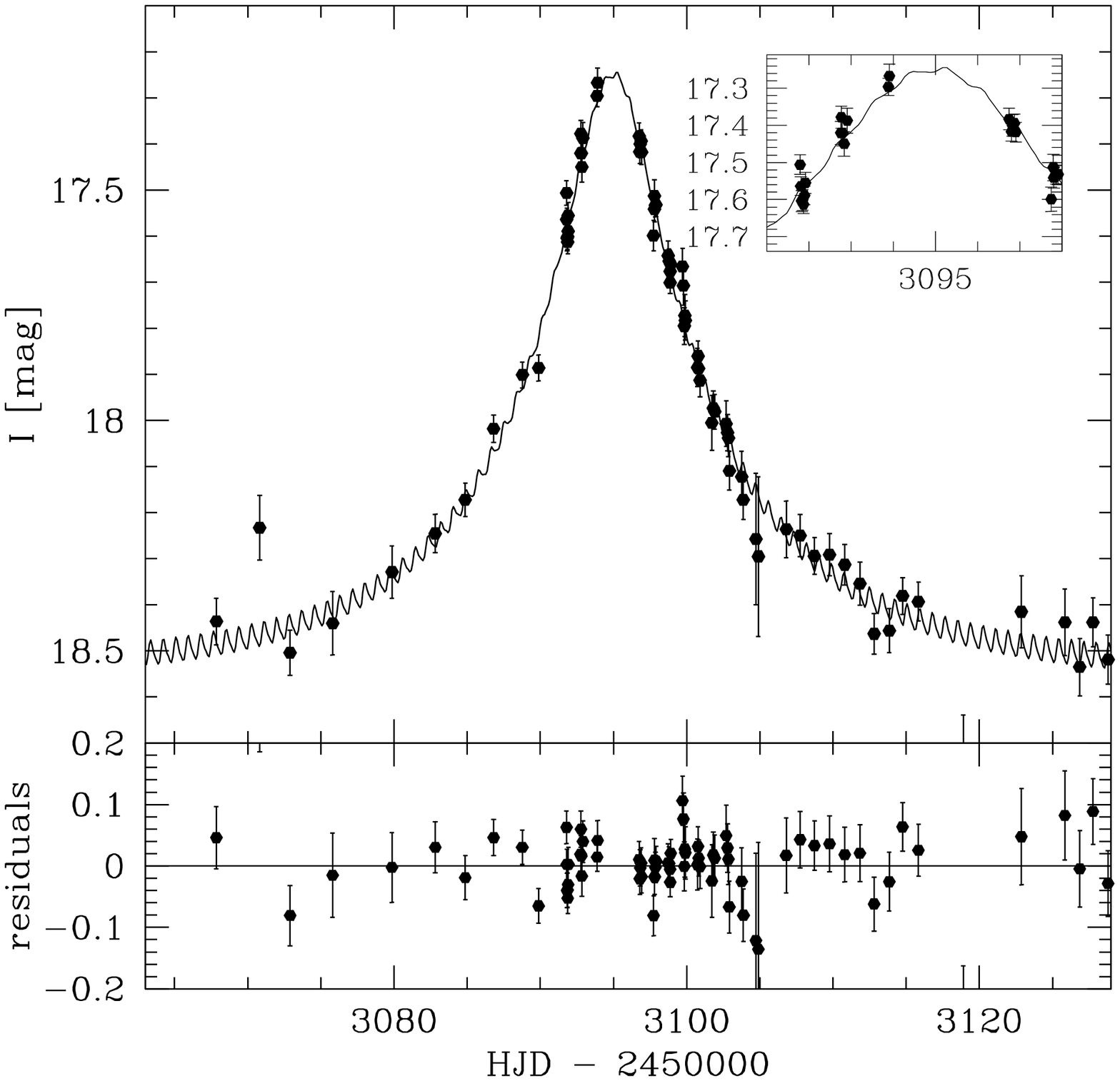}}
\FigCap{Data and best-fit microlensing models with variable baseline of
event OGLE175907.77--305519.1, The solid line shows the
variable blend model.}
\end{figure} 
\begin{figure}[htb]
\centerline{\includegraphics[width=9.3cm]{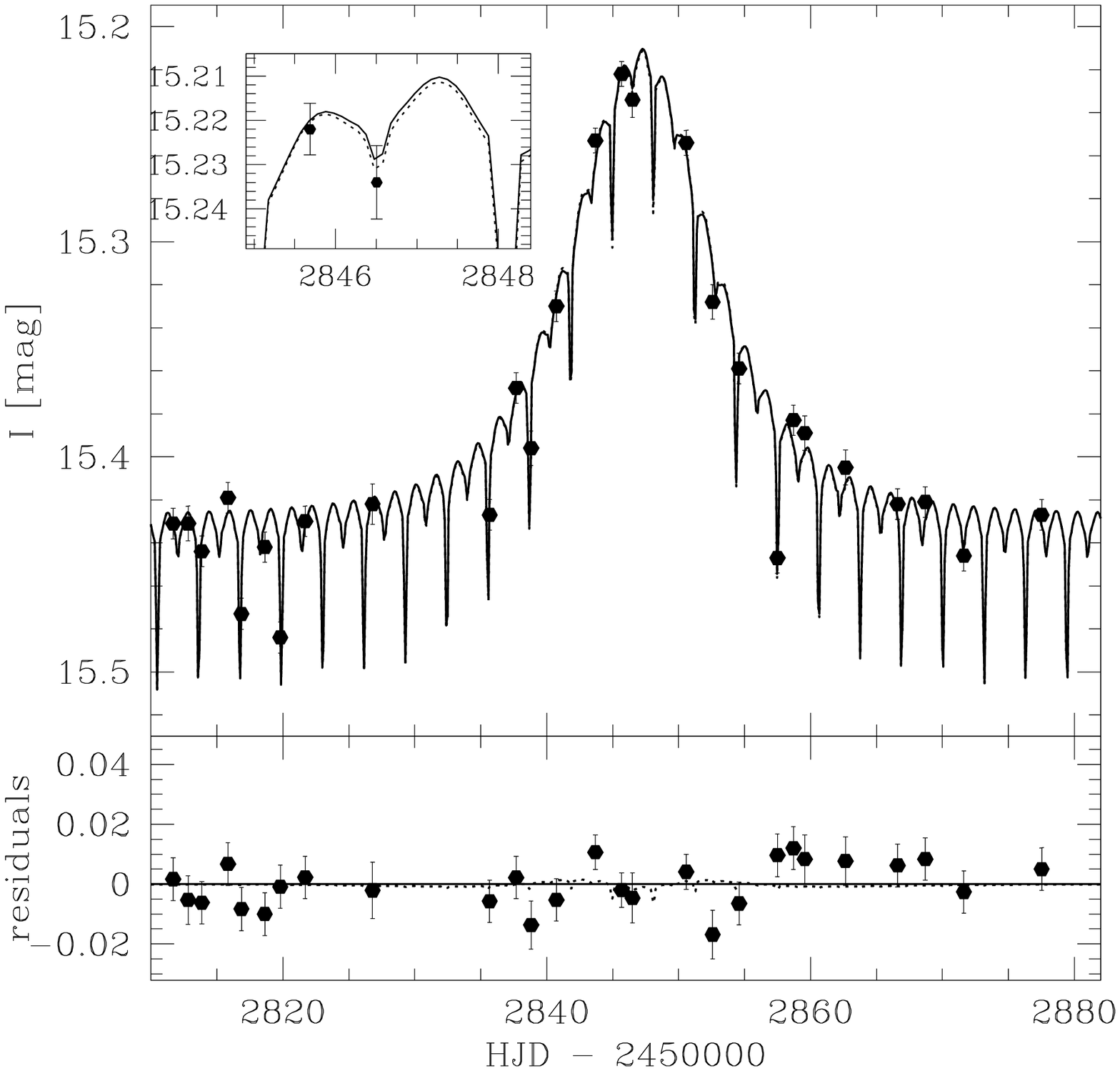}}
\FigCap{Data and best-fit microlensing models with variable baseline of
event OGLE180047.11--285934.5. The dotted and solid lines are for the
variable source and variable blend models, respectively.}
\end{figure}

In the case of OGLE180047.11--285934.5 with eclipsing baseline (Fig.~7)
both the variable source and variable blend models can fit the data with
similar accuracy. We are unable to distinguish these two models mainly
because the microlensing amplification for this event is very small,
which causes the difference in the depths of amplified eclipses in these
two models to be also small. The inset in Fig.~7 shows one data point,
which occurred during the secondary eclipse almost at the peak of the
event. However, this point is in agreement with both the variable source
and variable blend models.

Another event with eclipsing baseline, OGLE175535.13--331533.3 was not
fitted with the standard microlensing model with variable baseline, because
its light curve exhibits both parallax and binary source effects. This
event will be described and modeled in Wyrzykowski \etal (2006, in
preparation).

The remaining events were not fitted with the variable baseline
microlensing event model due to an insufficient number of data points
collected during the events. The sparse sampling allows a variety of models
to fit the data and the obtained parameters have large error bars.

Interestingly, in the baseline of the event OGLE175708.46--302007.1 we
found another, weaker period. This object is probably a blend of two
variables, of which the one with weaker and shorter period seems to be
microlensed. However, again we were not able to find a unique model to the
data.

Two out of 21 events with periodic baseline, namely OGLE175618.55--294252.9
and OGLE180540.47--273427.5, shown in Fig.~8, are the most likely not
caused by microlensing.
\begin{figure}[htb]
\centerline{\includegraphics[width=11.7cm]{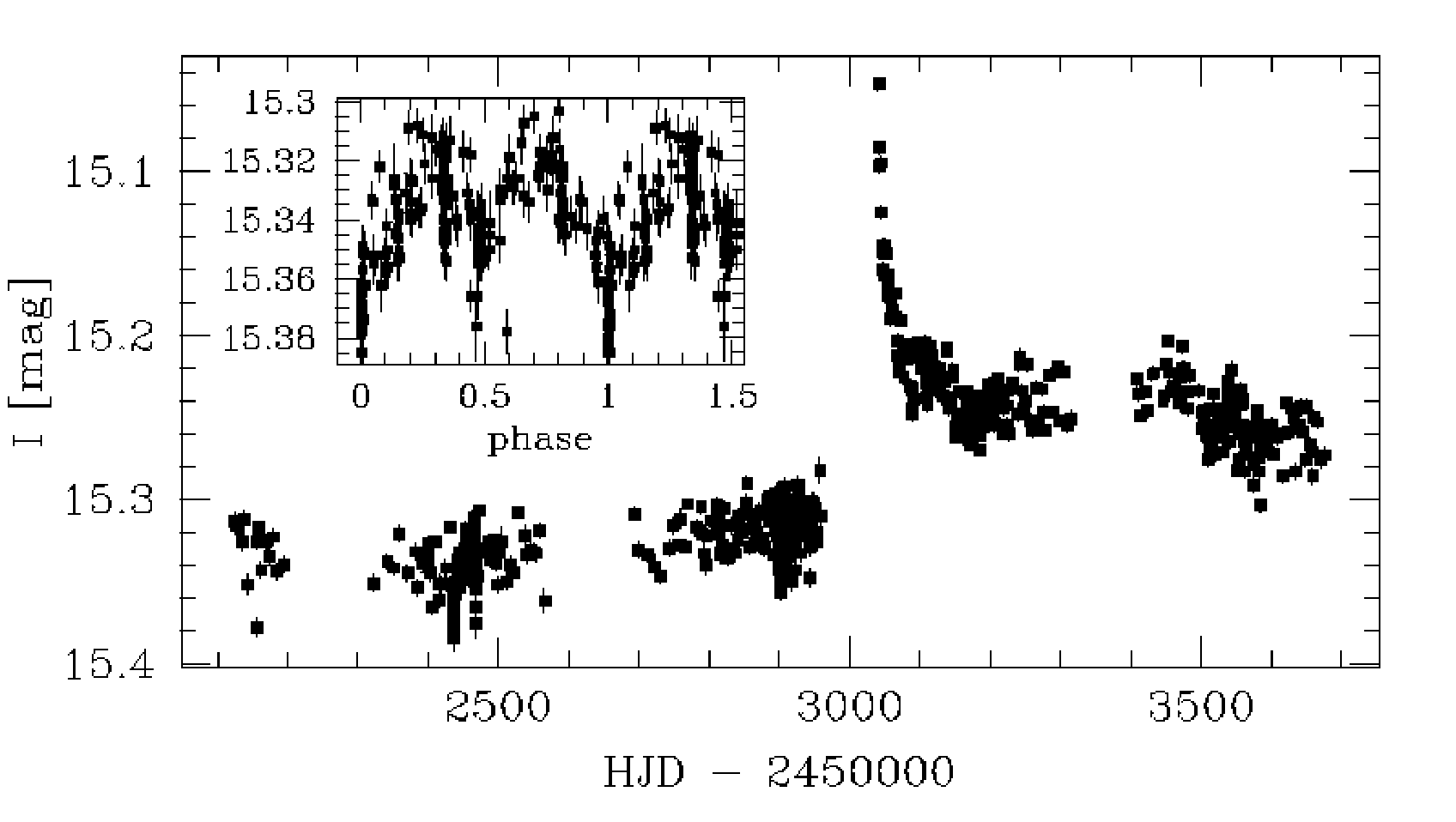}}

\centerline{\includegraphics[width=11.7cm]{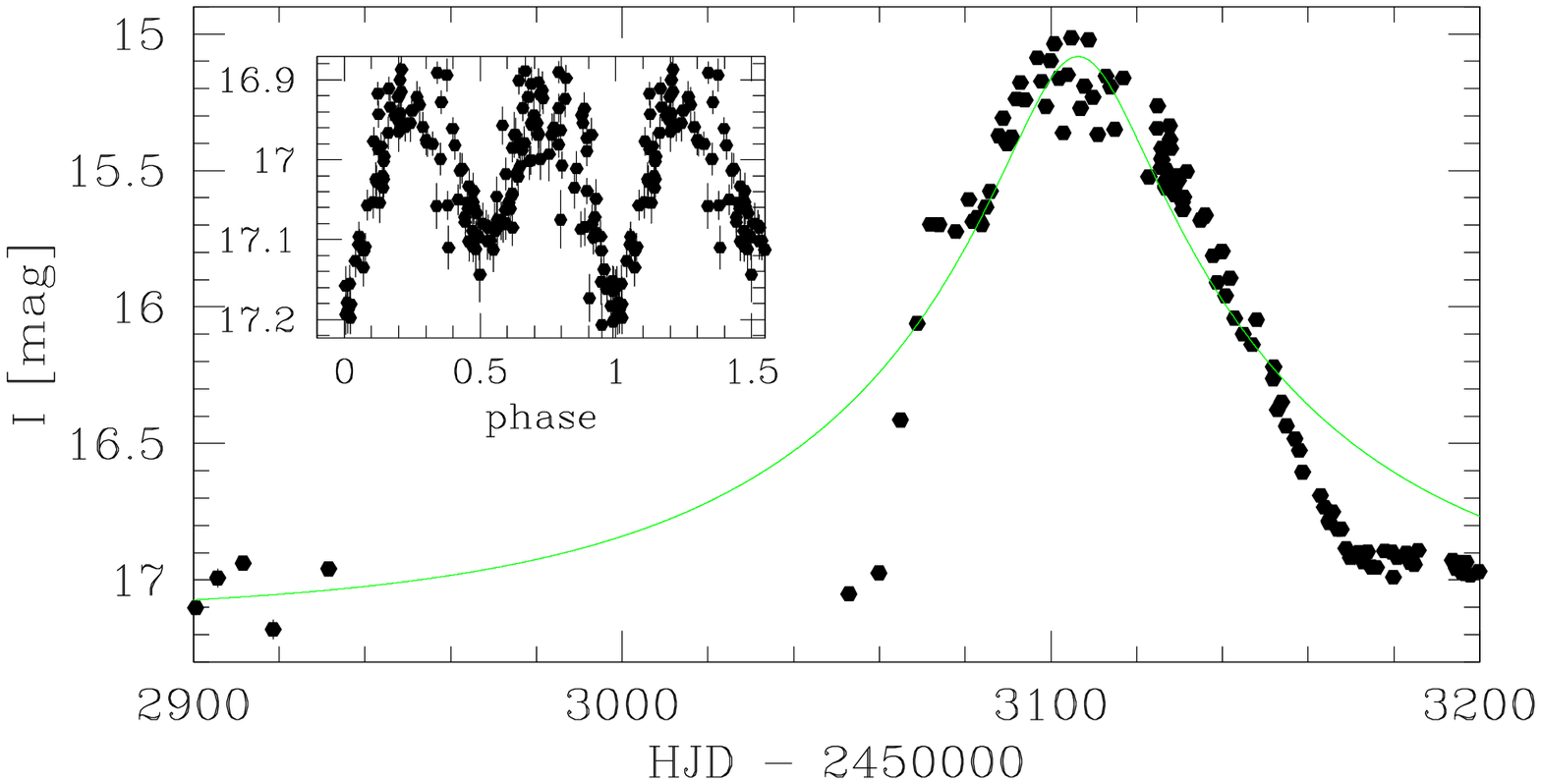}}
\FigCap{Light curves of OGLE175618.55--294252.9 ({\it top}) and
OGLE180540.47--273427.5 ({\it bottom}, with the best-fit standard
model). The insets show their baselines phase-folded with a period of
21.02317 and 3.96628 days, respectively. The first event may be a binary
lensing event, where the second caustic crossing was missed or has not
happened yet, or it is an eruptive variable star. The nature of the second
event is unclear. }
\end{figure}

The light curve of the first one resembles binary microlensing event, which
entered the caustic, however additional data collected in 2005 and at the
beginning of 2006 (also presented in Fig.~8) show no signature of the
second caustic crossing. Either it was missed in the gap between seasons,
not happened yet or the event is not microlensing but caused by some other
kind of variability, \eg a cataclysmic variable (CV) outburst. However,
both hypotheses make this event unusual.  Binary lens event would allow
determination of the lens distance directly from the analysis of the
baseline, assuming the observed variability comes from the lens. If this is
a CV outburst the system would have very long period (21.02317 days) among
CVs: the vast majority of CVs in the catalog of cataclysmic variables,
CVCAT (Downes \etal 2001), have periods shorter than 5 days.

The shape of the second unusual event, OGLE180540.47-273427.5, cannot be
explained with any known microlensing model. Its baseline folded with the
period of 3.96628 days suggests a contact binary variable star and thus
makes the CV outburst hypothesis plausible. However, again the period of
variability is very long compared to known CVs and the shape of the
brightening episode does not resemble that of CVs.

Both events require further analysis and additional follow-up photometric
and spectroscopic observations to unveil their real nature.
\vspace*{11pt}
\Section{Irregular Baseline Events}
\vspace*{9pt}
Table~4 lists 111 candidates for microlensing events with irregular
variability in the baseline: their name, mean baseline brightness in {\it
I}-band and cross-identification with events from the EWS, wherever
possible. Appendix~B presents the light curves of these events. All the
light curves are shown for the time span of ${\rm 2100<HJD{-}2450000<3350}$
days. The minimum and maximum magnitudes of each event is shown in the
upper (left and right, respectively) part of each panel.
\renewcommand{\arraystretch}{1.1}
\MakeTableSep{lcc|lcc}{12.5cm}{Microlensing event candidates with irregular
baseline. At the bottom five possible eclipsing baseline events are
listed.}
{\hline
\multicolumn{1}{c}{name} & \uprule $\langle I_{\rm base}\rangle$ & EWS & \multicolumn{1}{c}{name} & $\langle I_{\rm base}\rangle$ & EWS \\
\multicolumn{1}{c}{OGLE-} & \dorule       [mag]              &     & \multicolumn{1}{c}{OGLE-}  &         [mag]             & \\
\hline
\uprule
173511.27-265652.5 & 16.22 & 2002-BLG-221 & 175559.20-284555.3 & 18.57 & 2002-BLG-076 \\
173649.56-270326.5 & 17.07 & - & 175602.67-294724.3 & 13.52 & - \\
173841.53-211137.1 & 15.04 & 2002-BLG-241 & 175603.52-335323.9 & 15.23 & 2003-BLG-311 \\
173855.97-211617.9 & 19.49 & - & 175603.95-312649.7 & 16.21 & - \\
174335.42-335711.5 & 17.63 & 2004-BLG-437 & 175623.74-304338.0 & 16.05 & - \\
174346.78-342717.2 & 17.54 & - & 175624.81-323804.4 & 14.43 & - \\
174422.49-334729.3 & 16.72 & 2003-BLG-212 & 175626.92-290430.1 & 15.03 & - \\
174432.86-235842.4 & 15.74 & - & 175627.18-333626.0 & 15.30 & - \\
174501.73-333707.2 & 13.95 & - & 175627.59-284211.8 & 16.21 & 2003-BLG-442 \\
174522.43-333703.7 & 16.97 & - & 175631.70-304949.9 & 15.29 & - \\
174635.41-334619.7 & 14.76 & 2004-BLG-361 & 175641.98-324243.6 & 14.28 & 2004-BLG-296 \\
174635.42-334619.7 & 14.80 & - & 175657.24-330302.4 & 15.17 & - \\
174635.76-332510.7 & 17.04 & 2003-BLG-010 & 175712.40-305041.4 & 16.39 & - \\
174639.69-343542.6 & 14.08 & - & 175717.60-293513.2 & 13.90 & 2002-BLG-290 \\
174657.43-331937.4 & 15.53 & - & 175743.18-285911.3 & 18.24 & - \\
174732.74-340115.4 & 15.82 & - & 175746.93-295656.0 & 14.78 & 2004-BLG-077 \\
174737.69-332338.1 & 16.62 & 2002-BLG-019 & 175747.82-300450.1 & 13.87 & - \\
174825.61-232313.4 & 15.22 & 2003-BLG-430 & 175818.81-294918.7 & 15.18 & - \\
174826.62-343333.0 & 14.57 & - & 175834.71-295834.3 & 15.88 & 2004-BLG-026 \\
174904.99-345252.3 & 13.12 & - & 175846.00-290743.3 & 15.47 & 2004-BLG-029 \\
174921.09-352558.8 & 14.60 & - & 175903.04-271119.8 & 15.57 & 2003-BLG-095 \\
174938.25-223143.1 & 14.74 & - & 175912.23-295324.2 & 17.08 & - \\
174951.98-303628.8 & 15.42 & - & 175919.19-295316.1 & 14.64 & 2004-BLG-580 \\
174953.66-300640.8 & 16.08 & 2004-BLG-009 & 175941.94-283506.7 & 15.62 & - \\
174954.62-351932.1 & 15.20 & - & 180024.58-313821.2 & 15.56 & - \\
175020.91-342856.1 & 15.88 & - & 180029.65-291209.7 & 17.63 & 2002-BLG-356 \\
175047.59-293005.0 & 16.34 & 2003-BLG-120 & 180032.79-315100.3 & 17.63 & 2003-BLG-389 \\
175111.06-294754.1 & 14.21 & - & 180035.39-322702.6 & 15.43 & 2003-BLG-088 \\
175120.74-302655.6 & 14.02 & - & 180051.26-311244.1 & 15.62 & 2003-BLG-035 \\
175121.12-294109.3 & 18.28 & - & 180112.43-294812.0 & 13.42 & - \\
175125.38-322147.6 & 15.77 & - & 180129.50-302841.5 & 18.00 & - \\
175130.07-290810.3 & 15.85 & - & 180149.88-312136.2 & 18.66 & 2003-BLG-034 \\
175134.32-295519.3 & 16.76 & 2002-BLG-193 & 180209.11-301101.4 & 15.64 & 2003-BLG-135 \\
175206.25-292116.6 & 15.19 & - & 180215.81-320242.2 & 18.10 & 2004-BLG-145 \\
175221.14-333426.9 & 13.89 & - & 180308.17-314411.6 & 16.04 & 2002-BLG-034 \\
175226.40-292432.0 & 14.35 & - & 180314.20-273106.6 & 15.33 & - \\
175235.04-334343.1 & 15.86 & - & 180348.19-293301.8 & 15.69 & - \\
175239.24-290150.1 & 16.69 & 2004-BLG-362 & 180359.54-314950.2 & 14.28 & - \\
\dorule 175256.12-294509.2 & 17.59 & - & 180423.57-285120.0 & 13.29 & - \\
\hline}

\setcounter{table}{3}
\MakeTable{lcc|lcc}{12.5cm}{Concluded}
{\hline
\multicolumn{1}{c}{name} & \uprule $\langle I_{\rm base}\rangle$ & EWS & \multicolumn{1}{c}{name} & $\langle I_{\rm base}\rangle$ & EWS \\
\multicolumn{1}{c}{OGLE-} & \dorule       [mag]              &     & \multicolumn{1}{c}{OGLE-} &         [mag]             & \\
\hline
\uprule
175257.97-300626.3 & 18.11 & - & 180424.07-315220.7 & 16.95 & 2002-BLG-137 \\
175308.43-304353.9 & 14.61 & - & 180424.59-275340.7 & 16.47 & - \\
175341.55-303119.0 & 16.65 & - & 180429.93-290550.3 & 17.61 & - \\
175350.55-293117.7 & 14.18 & - & 180507.09-274309.2 & 16.63 & - \\
175355.20-291021.5 & 17.42 & 2002-BLG-041 & 180542.81-323020.2 & 18.20 & - \\
175405.62-314555.2 & 16.25 & 2004-BLG-018 & 180613.39-283613.4 & 18.58 & - \\
175409.89-285636.9 & 17.36 & - & 180623.05-303612.4 & 18.71 & - \\
175417.90-295058.1 & 14.27 & - & 180625.20-310346.4 & 16.39 & 2003-BLG-171 \\
175422.81-302324.5 & 14.98 & 2004-BLG-600 & 180628.66-273742.9 & 13.38 & 2004-BLG-383 \\
175425.39-355519.5 & 15.12 & - & 180638.79-274702.3 & 13.13 & - \\
175435.35-291929.0 & 15.93 & - & 180700.89-274548.4 & 15.93 & - \\
175440.01-315035.3 & 16.32 & - & 180903.23-255755.1 & 15.12 & - \\
175453.32-304307.4 & 15.07 & - & 180926.89-270716.2 & 14.87 & 2003-BLG-133 \\
175455.98-290113.8 & 16.57 & - & 180933.92-292018.8 & 15.26 & - \\
175456.39-295441.5 & 13.95 & 2004-BLG-290 & 181054.57-272434.9 & 16.48 & 2004-BLG-051 \\
175459.82-335325.9 & 18.23 & - & 181332.21-281254.9 & 15.43 & - \\
175500.72-310411.0 & 18.76 & - & \\ & & \\
\hline
\uprule
173551.54-271144.9 & 14.67 & 2002-BLG-220   &  175606.17-293334.4  & 13.99
       & - \\
175206.95-340704.2 & 16.18 & -   & 175933.80-300000.6 & 15.74 & - \\
\dorule 175320.16-293003.7 & 16.17 & 2002-BLG-363  & & & \\
\hline}

There is a large variety of variability types in the baselines of presented
events: from almost constant with some sporadic fluctuations to clear,
large amplitude variability. The shape of most events with bright baselines
($\langle I\rangle\leq16$~mag) resemble the OGLE Small Amplitude Red
Giants (OSARGs, Wray, Eyer and Paczy{\'n}ski 2004), \eg
OGLE174953.66--300640.8. The nature of the remaining variable baselines is
not clear at the moment and requires further multicolor or spectroscopic
observations.

Brightening episodes of the vast majority of the events resemble standard
single lens microlensing, but there are also events which clearly exhibit
exotic behavior, \eg OGLE175912.23--295324.2 or OGLE180209.11--301101.4,
caused probably by a binary lensing system (\eg Mao and Paczy\'nski 1991).

In the analyzed dataset, in addition to 21 with periodic and 111 with
irregular baselines, we found 5 events with several data points
significantly below their mean baseline brightness. These events are listed
at the bottom of Table~4 and their light curves are shown in the last five
panels in Appendix~B. Some of the outlying data points are likely caused by
instrumental effects or bad weather, however some can be signatures of
eclipses. As their number is usually two or three it is impossible to
determine any periodicity. In order to detect and confirm eclipses more
observations or dedicated follow-up is necessary. Events with eclipsing
variability in the baseline are one of the most valuable, as their
baselines can be used for determining information about the lens and the
source. Two of these events are particularly suspected to be in fact
binary source microlensing events: OGLE173551.54--271144.9 and
OGLE175206.95--340704.2, as their light curves exhibit some small
systematic deviations from the standard model during the magnification.
Such events may in some cases lead to a unique determination of the lens
mass (Wyrzykowski \etal 2006, in preparation).

\Section{Variable Events Warning System}
Most of the events presented in this paper with periodic baselines were not
fitted with any model due to the small number of observations taken during
the magnification. These cases are unfortunately lost. However, in order to
make good use of forthcoming events it is necessary to detect them in the
early stage of microlensing and then follow them up, especially during the
highest magnification.

Here we briefly describe a system, called Variable Events Warning System
(VEWS), a preliminary version of which is already working as an extension
of the existing OGLE Early Warning System (EWS). Similar to the EWS
(Udalski 2003), the new system checks the incoming photometric data from
the telescope in order to detect possible microlensing brightenings. These
two systems are complementary as the EWS investigates only constant stars,
while the VEWS checks variable stars. As mentioned above, the most
valuable variable baseline events are those with eclipsing baseline
variability, as they can be used to determine the lens or source distances
directly, as well as other parameters of the event. As a first attempt,
VEWS is based only on the catalog of eclipsing variables.  Currently the
VEWS uses about 7000 stars, selected from the catalog of about 10000
eclipsing binaries found by Devor (2005) in the OGLE-II Galactic bulge
data, which were cross-correlated with stars in 56 OGLE-III Bulge fields
frequently monitored since 2006. In the future the VEWS will be based on a
complete set of eclipsing variable stars in the OGLE-III Galactic bulge
data, when such a catalog becomes available.

As a result the system produces alerts of possible ongoing microlensing
events of eclipsing variable stars. Detailed information of any event and
the baseline eclipsing binary star will be provided to observers worldwide
and shown at the OGLE EWS website\footnote{{\it
http://ogle.astrouw.edu.pl/ogle3/ews/ews.html}} to allow photometric and
spectroscopic follow-up observations by other observatories.

The detection rate of events with periodic baselines is expected to be
around seven per year, with at least three on eclipsing variable stars.
This estimate is based on our detection of 13 new periodic baseline events,
not detected by the EWS (6 new eclipsing) in comparison with 1137 events
detected by the EWS in our 82 fields in years 2001--2004. Assuming a
typical OGLE detection rate of 600 regular microlensing events per year, we
expect half a dozen microlensing events with periodic variable baselines
per season.

\Section{Conclusions} 
In this paper we presented microlensing events with variable baselines and
showed that they are common in the Galactic bulge fields. Analysis of
variability amplitude behavior during gravitational microlensing
amplification may lead to the determination of parameters of the event,
removing common degeneracy due to blending. The change of the variability
amplitude depends on the blending parameter $f_s$, whose determination is
an important issue in almost all microlensing problems. For example, in the
optical depth determinations, an incorrect blending parameter may
dramatically change the timescale of an event and thus affect the estimated
optical depth (\eg Sumi \etal 2006).

In addition, events with variable baseline can also potentially be used for
the determination of physical parameters of the microlensing event, which
are not measurable for a standard microlensing event. Using characteristics
of a given variability type, \eg eclipsing binary, Cepheid or RR Lyrae, it
might be possible to obtain the distance to the source, if the variable
star acts as the lensed source. Another interesting possibility may occur
if the lensing object is a variable and acts as a blend. In such cases,
especially when there are binary lens signatures in the light curve and
eclipsing variability in its baseline, it might be possible to determine
the distance and even the mass of the lens, which was so far only possible
for very few events (\eg Alcock \etal 2001, An \etal 2002, Jiang \etal
2004).

In order to allow detailed follow-up observations of interesting variable
baseline events we are testing the Variable Events Warning System (VEWS),
which can detect microlensing brightening occurring on eclipsing binary
stars.

\Acknow{We would like to thank Profs. Bohdan Paczy{\'n}ski, Tsevi Mazeh
and Drs. Vasily Belokurov, Wyn Evans, Martin Smith, Omer Tamuz and Szymon
Koz{\l}owski for their continuous support for this project.

This paper was partially supported by the MNiSW BST grant to the Warsaw
University Observatory, NSF grant AST-0204908, NASA grant NAG5-12212 and by
the European Community's Sixth Framework Marie Curie Research Training
Network Programme, Contract No. MRTN-CT-2004-505183 "ANGLES". {\L}W also
acknowledges support from the visitor programme at the Jodrell Bank
Observatory.}

\newpage
\begin{figure}[p]
\begin{center}
{\bf Appendix A}
\vskip12pt
{\bf Microlensing events candidates with periodic variability in the
baseline}\\
The left column contains the whole light curve of the event and
its name while the right column shows the phase-folded baseline and period
in days.
\end{center}
\hglue-9mm{\includegraphics[width=14cm]{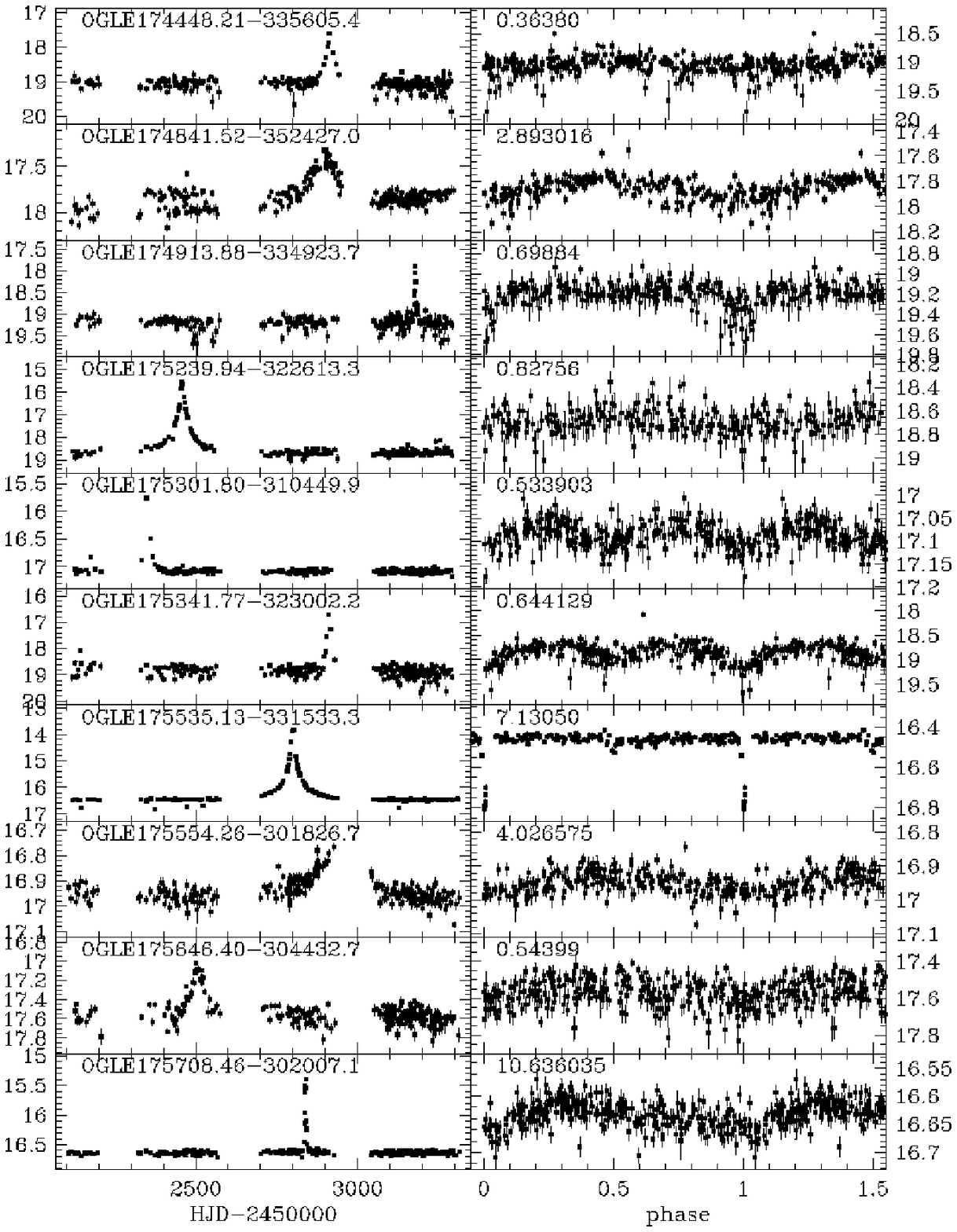}}
\end{figure}
\begin{figure}[htb]
\hglue-9mm{\includegraphics[width=14cm]{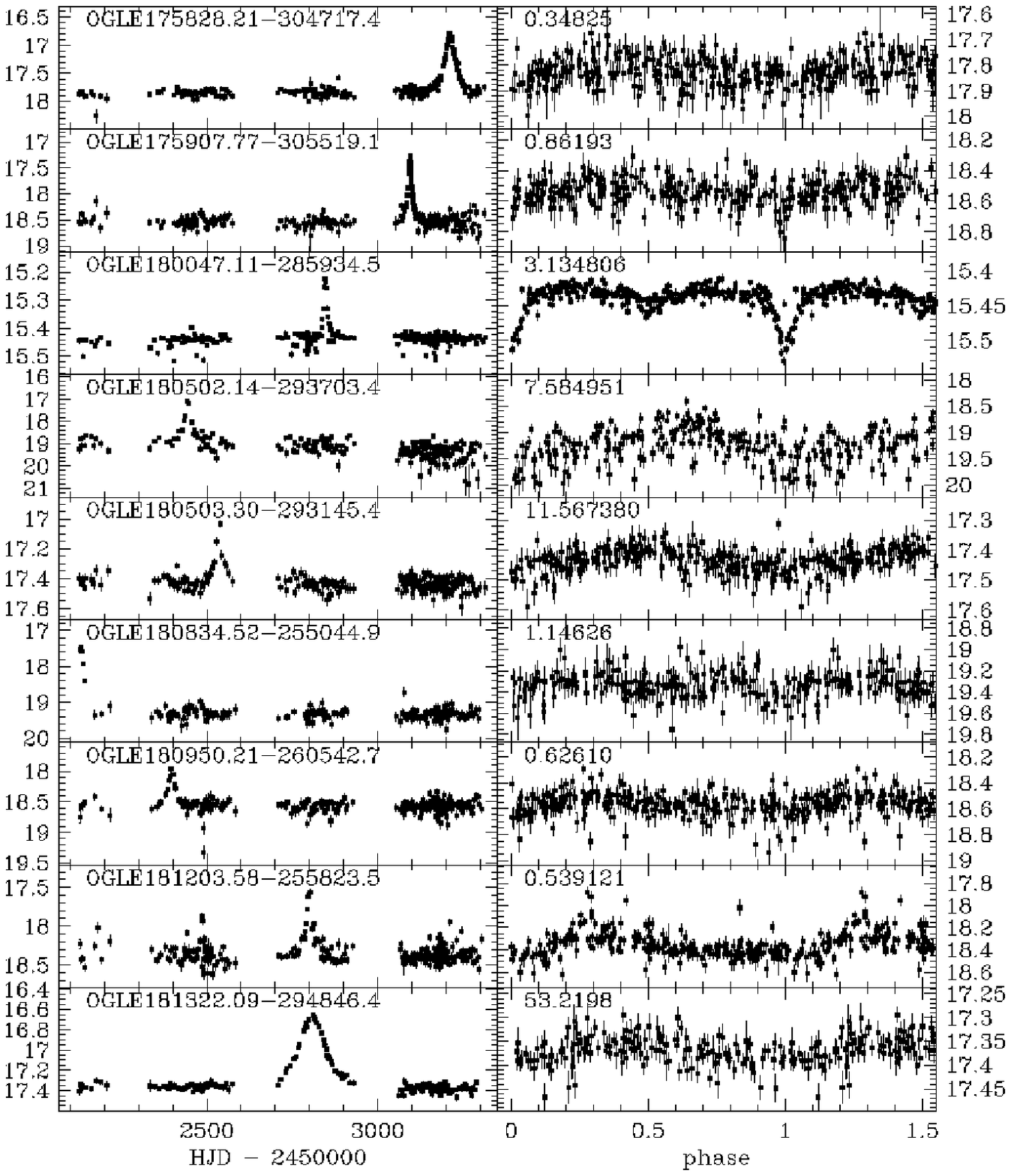}}
\end{figure}
\begin{figure}[p]
\begin{center}
{\bf Appendix B}
\vskip6pt
{\bf Candidates for microlensing events with irregular variability in the
baseline}\\ 
The horizontal scale is the HJD-2450000 spanning from 2100 to
3350 days. The numbers in each panel are the minimum ({\it left}) and
maximum ({\it right}) brightness in magnitudes in the {\it I}-band. The
last five events are potential eclipsing baseline events listed at the end
of Table~4.
\end{center}
\vspace*{-5pt}
\hglue-5mm{\includegraphics[width=14cm]{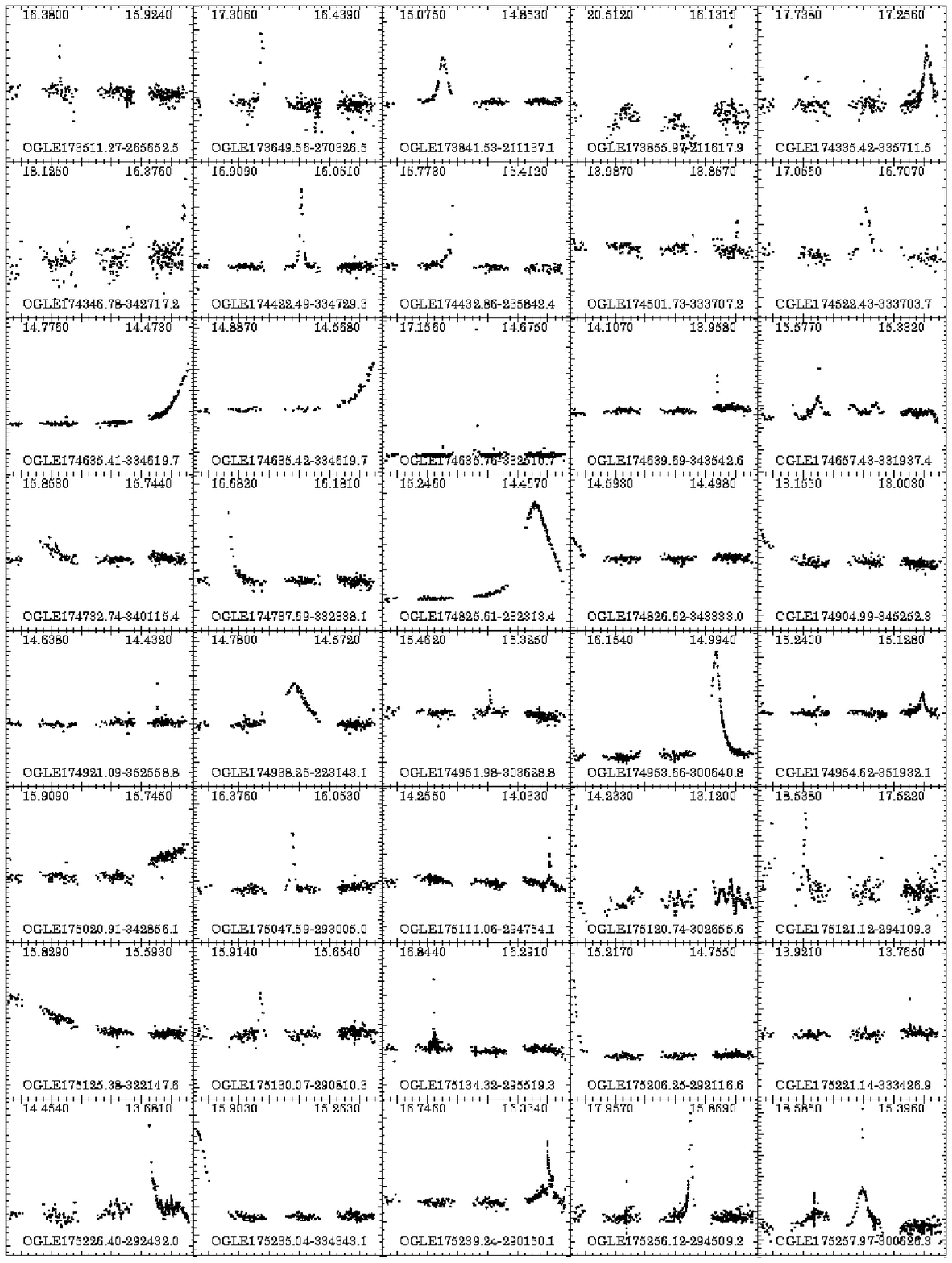}}
\end{figure}
\begin{figure}[p]
\hglue-5mm{\includegraphics[width=14cm]{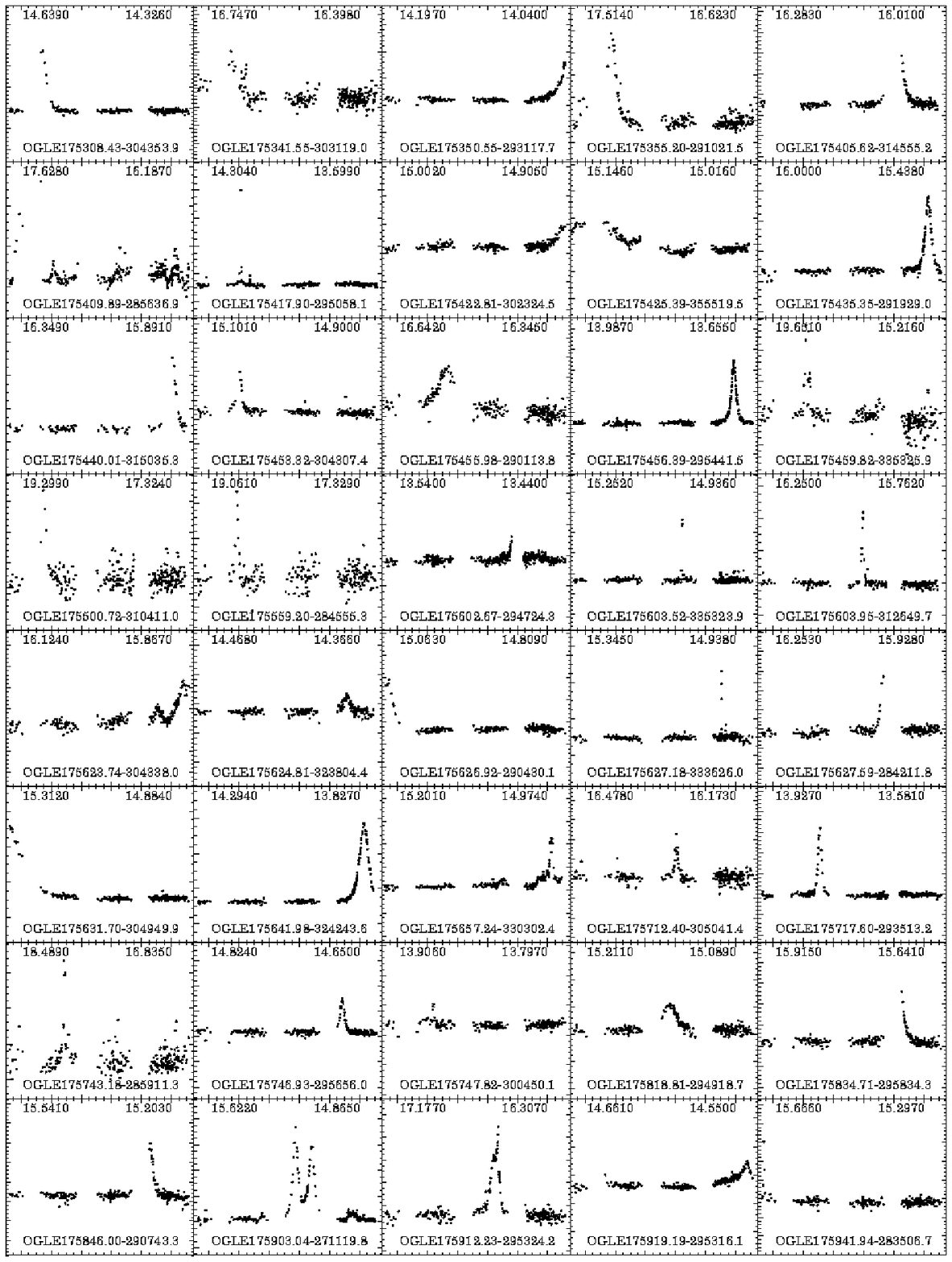}}
\end{figure}
\begin{figure}[p]
\hglue-5mm{\includegraphics[width=14cm]{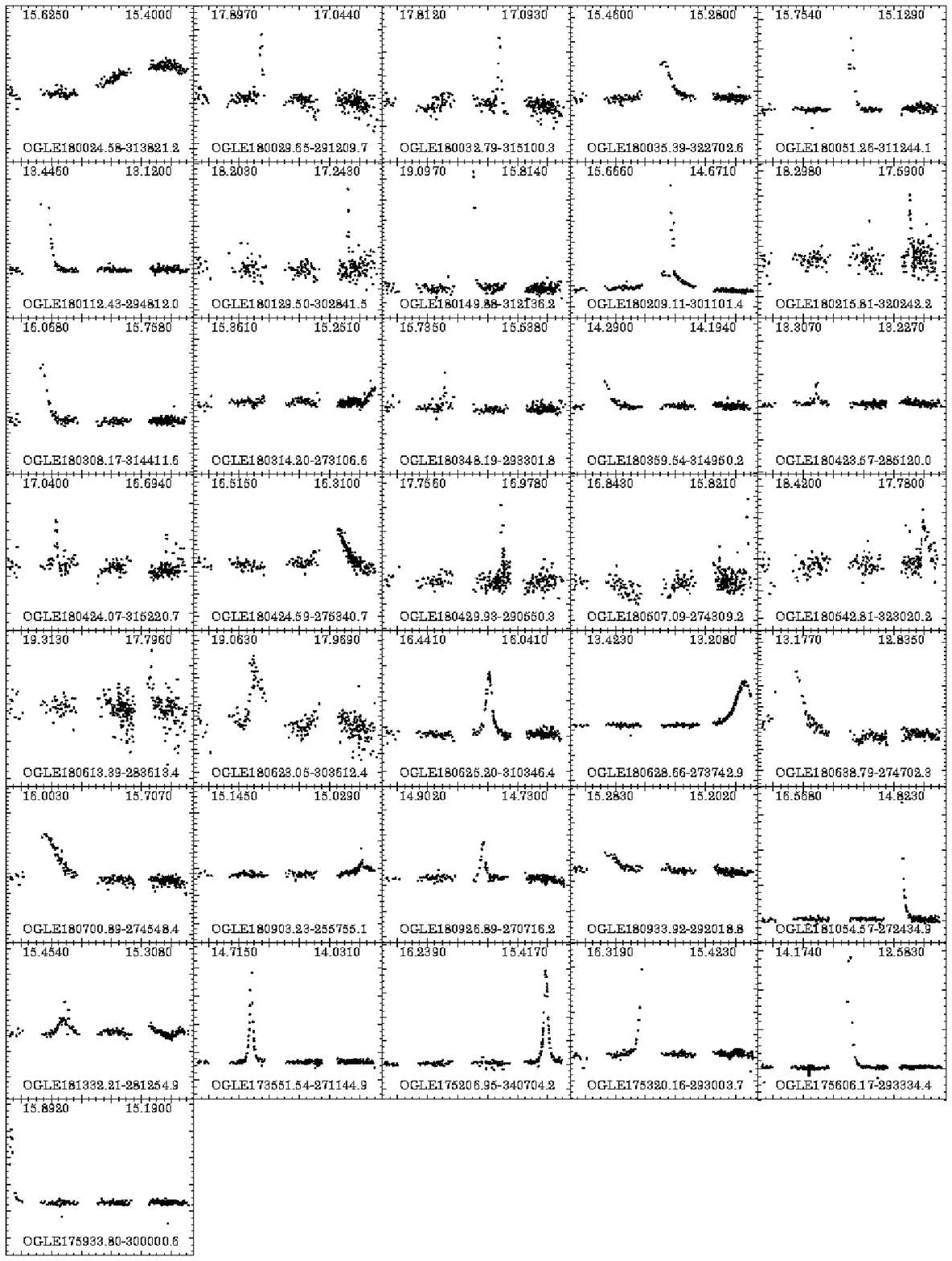}}
\end{figure}
\end{document}